\documentclass{imsart}

\RequirePackage{amsthm,amsmath,amsfonts,amssymb}
\RequirePackage{natbib}
\RequirePackage{bm}
\RequirePackage[colorlinks,citecolor=blue,urlcolor=blue,backref=page,backref=page]{hyperref}
\RequirePackage{graphicx}
\usepackage{algorithm}
\usepackage{algorithmicx}
\usepackage{algpseudocode}
\usepackage{multicol}
\usepackage{multirow}
\usepackage{subcaption}
\usepackage{comment}
\usepackage[english]{babel}

\pubyear{2025}
\arxiv{arXiv:2410.21603}
\volume{TBA}
\issue{TBA}
\firstpage{1}
\lastpage{1}

\startlocaldefs
\theoremstyle{plain}

\newtheorem{theorem}{Theorem}[section]

\theoremstyle{definition}

\usepackage[T1]{fontenc}
\usepackage{babel}
\usepackage[table]{xcolor}
\usepackage{collcell}
\usepackage{hhline}
\usepackage{pgf}
\usepackage{multirow}

\def\colorModel{hsb} 

\newcommand\ColCell[1]{
  \pgfmathparse{#1<50?1:0}  
    \ifnum\pgfmathresult=0\relax\color{white}\fi
  \pgfmathsetmacro\compA{0}      
  \pgfmathsetmacro\compB{#1/500} 
  \pgfmathsetmacro\compC{0.95}      
  \edef\x{\noexpand\centering\noexpand\cellcolor[\colorModel]{\compA,\compB,\compC}}\x #1
  } 
\newcolumntype{E}{>{\collectcell\ColCell}m{0.4cm}<{\endcollectcell}}  
\newcommand*\rot{\rotatebox{90}}

\usepackage[table]{xcolor}
\usepackage{array, hhline}
\usepackage{rotating} 

\renewcommand{\rot}[1]{\rotatebox{90}{#1}}

\newcommand{\colorednum}[1]{%
  \ifnum#1=0
    \hspace{1em}%
  \else
    \ifnum#1>90
      \textcolor{red}{#1}%
    \else
      \ifnum#1<10
        \textcolor{blue}{#1}%
      \else
        #1%
      \fi
    \fi
  \fi
}

\theoremstyle{remark}

\endlocaldefs

\begin{document}

\begin{frontmatter}
\title{Approximate Bayesian Computation with Statistical Distances for Model
Selection}
\runtitle{ABC-MC}

\begin{aug}
\author{\fnms{Clara} \snm{Grazian}\thanksref{addr1,addr2}\ead[label=e1]{clara.grazian@sydney.edu.au}}

\runauthor{Grazian}

\address[addr1]{Carslaw Building, University of Sydney, Camperdown Campus, Sydney, Australia \printead{e1} 
}
\address[addr2]{ARC Training Centre in Data Analytics for Resources and Environments (DARE), Sydney, Australia 
}
\end{aug}

\begin{abstract}
Model selection in the presence of intractable likelihoods remains a central challenge in Bayesian inference. Approximate Bayesian computation (ABC) provides a flexible likelihood-free framework, but its use for model choice is known to be sensitive to the choice of summary statistics, often leading to poorly calibrated posterior model probabilities. Recent ABC variants based on statistical distances allow comparisons to be performed directly on empirical distributions, avoiding data reduction and offering improved theoretical guarantees under suitable conditions. This paper provides a systematic evaluation of discrepancy-based ABC methods for Bayesian model selection, focusing on their empirical behavior across a range of simulation settings and levels of model complexity. We compare full data ABC approaches based on Wasserstein, Cram\'er–von Mises, and maximum mean discrepancy metrics with summary-statistic-based ABC and neural network classifiers. The results highlight settings in which full data ABC yields stable and well-calibrated posterior model probabilities, as well as scenarios where performance degrades due to model overlap or dependence. An application to toad movement models illustrates the practical implications of these findings. Overall, the study clarifies the strengths and limitations of discrepancy-based ABC for likelihood-free model choice and provides guidance for its use in realistic inferential settings.
\end{abstract}



\end{frontmatter}

\section{Introduction}\label{intro}

The selection of a model that best explains a dataset or process is central to statistics and relevant across disciplines. While no model can fully capture all nuances in complex phenomena, approximating the true data-generating process (DGP) can yield valuable insights \citep{molnar2022}. Importantly, the best model is not fixed; as new evidence arises, previously accepted models may require refinement or replacement \citep{gelman2013}.

Bayesian statistics provides a natural framework for model selection. Bayes' theorem enables the updating of prior beliefs about candidate models as new data becomes available. Each model's support is quantified through its posterior probability, i.e. the probability of the model given the data, enabling probabilistic model comparison. The model with the highest posterior probability can be selected, or model averaging can be used to weigh models based on these probabilities \citep{raftery1995, wasserman2000}.

However, many models involve complex interdependencies or latent variables that render their likelihood functions intractable \citep{martin2024}. In such cases, exact posterior probabilities cannot be computed, motivating likelihood-free methods such as approximate Bayesian computation (ABC). While ABC is typically used for parameter inference, it can also support model selection, provided all models can be simulated from. ABC estimates model support by weighting how often each model generates data that is ``close enough'' to the observed data, thus approximating its posterior probability.

While ABC has traditionally relied on low-dimensional summary statistics to mitigate the curse of dimensionality, this strategy is known to be problematic for model selection \citep{robert2011}. Recent work has proposed replacing summaries with statistical distances between empirical distributions, yielding ABC procedures that operate directly on the observed and simulated data \citep{park2015, bernton2019, nguyen2020, frazier2020}. Alternatively, \cite{forbes2022summary} have introduced a framework that constructs surrogate posteriors using finite Gaussian mixtures, employing inverse regression techniques and then utilizes distributional metrics to compare surrogate posteriors.  These discrepancy-based methods have been shown to possess attractive theoretical properties for parameter inference, including robustness to certain forms of model misspecification. Their implications for Bayesian model selection, however, remain less well understood, particularly in settings involving overlapping models or dependent data. 

Machine learning, particularly deep learning, has been used to improve ABC by automatically learning low-dimensional summaries \citep{blum2010non,sheehan2016deep,jiang2017learning}. While such approaches have shown promise for parameter inference, they typically require large simulated training sets and can be sensitive to model misspecification. Deep learning has also been applied to related classification tasks \citep{bures2023organic}, though these differ conceptually from likelihood-free Bayesian model selection, which compares generative models via simulation under each candidate model. Classifiers trained with cross-entropy loss can approximate posterior model probabilities, but they rely on discriminative approximations that may be poorly calibrated when models overlap. In contrast, ABC compares simulated and observed data directly under each model, providing a principled framework for likelihood-free model comparison, which we evaluate in this paper against both summary-based ABC and deep learning classifiers.

The remainder of the paper is structured as follows. Section \ref{sec:background} reviews Bayesian model selection and introduces ABC. Section \ref{sec:ABCmodelchoice} discusses the use of ABC for model choice, highlighting its challenges and introducing full data methods, with a theoretical justification focused on the Wasserstein distance. Section \ref{sec:simu} presents a simulation study comparing full data, summary-statistics-based ABC approaches, and deep learning alternatives. Section \ref{sec:real} applies these methods to the toad movement models of \cite{marchand2017}. Finally, Section \ref{sec:conclu} concludes the paper. The code to replicate the results in this work is available at \url{https://github.com/bayesgra/ABC-MC}. 

\section{Background}
\label{sec:background}

\subsection{Notation and Setting}

Let $\bm{y} = (y_1, ..., y_n)^T \in \mathcal{Y}^n \subseteq \mathbb{R}^n$ denote the observed data, assumed to be generated from an unknown true distribution $P_{\bm{\theta}_0}^{(n)}$ within a model family $\mathcal{P} = \{P_{\bm{\theta}}^{(n)} : \bm{\theta} \in \Theta \subseteq \mathbb{R}^{d}\}$. The assumed model $M$ in this family is parameterized by $\bm{\theta}$ with prior distribution $\pi(\bm{\theta})$, yielding a likelihood function $p(\bm{y} | \bm{\theta})$ and posterior distribution $\pi(\bm{\theta} | \bm{y})$. The marginal likelihood is defined as $p(\bm{y}) = \int_\Theta p(\bm{y} | \bm{\theta}) \pi(\bm{\theta}) d\bm{\theta}$.

When the prior beliefs instead concerns a set of $K$ candidate models $M_1,\ldots, M_K$, each model $M_k$ has parameters $\bm{\theta}_k \in \Theta_k \subseteq \mathbb{R}^{d_k}$ with prior distribution $\pi_k(\bm{\theta}_k)$ and likelihood function $p_k(\bm{y} | \bm{\theta}_k) = p(\bm{y} | \bm{\theta}_k, M_k)$. Models are assigned prior probabilities $\pi(M_k) = \pi(M=M_k)$ such that $\sum\limits_{k=1}^K \pi(M_k) = 1$. The posterior probability for model $M_k$ is:
$$\pi(M_k | \bm{y}) = \frac{p_k(\bm{y}) \pi(M_k)}{\sum_{j = 1}^K p_j(\bm{y}) \pi(M_j)} \;\;\; \forall \; k = 1, ..., K,$$
where $p_k(\bm{y})$ is the marginal likelihood under model $M_k$. The true DGP $P_{M_0,\bm{\theta}_0}^{(n)}$ is assumed to belong to the broader set $\mathcal{P} = \{P_{M_k, \bm{\theta}_k}^{(n)} \; | \; \bm{\theta}_k \in \Theta_k \subseteq \mathbb{R}^{d_k}, k \in \{1, \ldots, K\} \}$.

To compare two models $M_i$ and $M_j$, the Bayes Factor is often used:
$$B_{ij}(\bm{y}) = \frac{p_i(\bm{y})}{p_j(\bm{y})} = \frac{\pi(M_i | \bm{y})}{\pi(M_j | \bm{y})} \bigg{/} \frac{\pi(M_i)}{\pi(M_j)}.$$

\subsection{Approximate Bayesian Computation}
\label{sub:ABC}

Obtaining posterior distributions for model selection is often analytically intractable. Even when likelihood functions are available, Markov chain Monte Carlo (MCMC) methods for posterior estimation and model comparison via Bayes factors can be computationally demanding and require careful tuning \citep{han2001markov}. These challenges are exacerbated in complex systems, where likelihoods may be unavailable or prohibitively expensive to evaluate, and where standard MCMC methods face scalability issues in large or high-dimensional settings \citep{bardenet2015}. 

ABC addresses this challenge by approximating posterior distributions in settings where the likelihood $p(\bm{y} | \bm{\theta})$ is unavailable. In its simplest form, rejection ABC generates parameters from the prior, simulates data from the corresponding generative model, and retains draws that produce simulated data sufficiently close to the observed data under a chosen discrepancy \citep{pritchard1999}. The resulting approximation targets a distribution of the form $\pi_\varepsilon (\bm{\theta} | \bm{\eta}(\bm{y}))$, which depends on user-specified summary statistics $\bm{\eta}(\cdot)$ and tolerance levels $\varepsilon$. Algorithm 1 in Appendix~A provides a detailed description of this version of ABC for parameter inference. Numerous extensions have been proposed to improve computational efficiency and sampling performance \citep{beaumont2002, marjoram2003, sisson2007, beaumont2009, delmoral2011}.

The choice of low-dimensional, informative summary statistics is known to be critical for ABC performance. In particular, \cite{barber2015} show that the mean squared error of ABC estimators deteriorates rapidly with the dimension $	q$ of the summary statistic, scaling as $\mathcal{O}(N^{-4/(q+4)})$. While a variety of approaches have been proposed to construct informative summaries \citep{joyce2008,nunes2010,fearnhead2012,
drovandi2011,gleim2013,drovandi2015,
martin2019}, their effectiveness for model selection is problem dependent and difficult to assess in practice.

Full data ABC methods compare empirical distributions of the observed and simulated data directly using statistical distances. In this framework \citep{drovandi2022}, the distance metric $\rho(\cdot, \cdot)$ is replaced by a discrepancy measure $\mathcal{D}(\cdot, \cdot)$ applied to empirical measures $\mu_{\bm{\theta}_0}$ and $\mu_{\bm{\theta}^*}$, constructed from observed and simulated samples, respectively. Here, $\mu_{\bm{\theta}_0}$ denotes the measure associated with the true distribution $P_{\bm{\theta}_0}^{(n)}$, while $\mu_{\bm{\theta}^*}$ denotes the measure associated with the simulated distribution $P_{\bm{\theta}^*}^{(n)} \in \mathcal{P}$. Since $P_{\bm{\theta}_0}^{(n)}$ is unavailable, empirical measures based on the observed and simulated data are used, namely $\hat{\mu}_{\bm{\theta}_0} = n^{-1} \sum\limits_{i=1}^n \delta_{y_i}$ and $\hat{\mu}_{\bm{\theta}^*} = n^{-1} \sum\limits_{i=1}^n \delta_{z_i}$ respectively, where $\delta_x$ denotes the Dirac measure at $x \in \mathcal{Y}$. Details are provided in Appendix~A (see Algorithm 2).

\cite{legramanti2023} provide theoretical justification for these methods using integral probability semimetrics (IPS), such as the Wasserstein distance or maximum mean discrepancy (MMD), and Rademacher complexity, establishing uniform concentration bounds for ABC posterior distributions, even under model misspecification and non-i.i.d. data. Empirical work by \cite{drovandi2022} shows that full data ABC can match or outperform summary-based ABC in parameter inference, with different discrepancies exhibiting problem-dependent behavior. In this paper, we examine the implications of discrepancy-based ABC methods for Bayesian model selection.

\section{Approximate Bayesian Computation for Model Selection}
\label{sec:ABCmodelchoice}

In Bayesian model selection, ABC approximates posterior model probabilities $\pi(M_k | \bm{y})$ through acceptance rates under each candidate model $M_k$, based on simulated data $\bm{z} \sim P_{M_k, \bm{\theta}_k}^{(n)}$ with likelihood $p_k(\bm{z} | \bm{\theta}_k)$. The ABC model choice (ABC-MC) framework, introduced by \citet{grelaud2009} in the context of Gibbs random fields and later extended to sequential Monte Carlo implementations by \cite{didelot2011}, has since been applied across a range of applied domains (see Algorithm 3 in Appendix~A).

However, model choice via ABC introduces challenges beyond standard inference. \citet{robert2011} showed that even with sufficient statistics and vanishing tolerance $\varepsilon \to 0$, the ABC Bayes factor between two models $B_{ij}^\eta(\bm{y})$, may fail to converge to the true Bayes factor $B_{ij}(\bm{y})$, due to the unbounded correction ratio
$$\frac{B_{ij}(\bm{y})}{B_{ij}^\eta(\bm{y})} = \frac{g_i(\bm{y})}{g_j(\bm{y})},$$
where $g_{\ell}(\bm{y})= p_{\ell}(\bm{y} | \bm{\theta}_{\ell}) / p_{\ell}^\eta(\bm{\eta}(\bm{y}) | \bm{\theta}_{\ell})$, with $p_{\ell}^\eta(\bm{\eta}(\bm{y}) | \bm{\theta}_{\ell})$ the likelihood function for the summary statistic under model $\ell=i,j$, quantifies the discrepancy between full and summary likelihoods. 

\citet{marin2013} derived sufficient conditions for summary statistics to yield consistent ABC model selection, requiring that only one candidate model be compatible with the observed summaries, a condition that is often difficult to satisfy in practice. While alternative strategies have been proposed to construct approximately sufficient summaries \citep{barnes2012,prangle2013}, such requirements highlight the intrinsic difficulty of summary-based ABC for model choice.

\subsection{ABC-MC with Statistical Distances}

In discrepancy-based ABC model selection, the ABC distance metric $\rho(\cdot, \cdot)$ is replaced with a discrepancy measure $\mathcal{D}(\cdot, \cdot)$ that operates on the probability measures of the observed data $\bm{y}$ and simulated data $\bm{z}$. We denote $\mu_{0,\bm{\theta}_0}$ as the measure associated with the true distribution $P_{M_0,\bm{\theta}_0}^{(n)}$, and $\mu_{k^*, \bm{\theta}^*_{k^*}}$ as the measure associated with the simulated data's distribution $P_{M_{k^*}, \bm{\theta}_{k^*}}^{(n)} \in \mathcal{P}$. The empirical distributions of $\bm{y}$ and $\bm{z}$ replace the true measures, and are similarly defined as $\hat{\mu}_{0,\bm{\theta}_0} = n^{-1} \sum\limits_{i=1}^n \delta_{y_i}$ and $\hat{\mu}_{k^*, \bm{\theta}^*_{k^*}} = n^{-1} \sum\limits_{i=1}^n \delta_{z_i}$ respectively. This forms the basis of Algorithm~\ref{alg:full-data_abc-mc}.

\begin{algorithm}
 \caption{Discrepancy-based ABC-MC}\label{alg:full-data_abc-mc}
 \begin{algorithmic}[1]
 \State \textbf{Input:} Given a set of observations $\bm{y}=(y_1, \ldots, y_n)^T$, a set of possible models $\{M_1, M_2, \ldots, M_K\}$, a tolerance level $\varepsilon$, and a discrepancy metric $\mathcal{D}(\cdot, \cdot)$:
 \For{$i = 1, \dots, N$}
 \Repeat
     \State Generate $M_{k^*}$ from $\pi(M_k)$, $k=1, \ldots, K$.
     \State Generate $\bm{\theta}_{k^*}$ from $\pi_{k^*}(\bm{\theta}_{k^*})$.
     \State Generate $\bm{z} = (z_1, ..., z_n)^T$ from $P_{M_{k^*}, \bm{\theta}_{k^*}}^{(n)}$.
 \Until{$\mathcal{D}(\hat{\mu}_{0,\bm{\theta}_0}, \hat{\mu}_{k^*, \bm{\theta}_{k^*}}) \leq \varepsilon$},
 \State Set $M^{(i)} = M_{k^*}$ and $\bm{\theta}^{(i)} = \bm{\theta}_{k^*}$.
 \EndFor
 \State \textbf{Output:} A set of values $(M^{(1)}, \ldots, M^{(N)})$ and $(\bm{\theta}^{(1)}, \ldots, \bm{\theta}^{(N)})$ from $\pi_{\varepsilon}(M_k|\bm{y})$ and $\pi_{\varepsilon,k} (\bm{\theta}_k | \bm{y},M_k)$, for $k=1, \ldots, K$, respectively.
 \end{algorithmic}
\end{algorithm}

Algorithm \ref{alg:full-data_abc-mc} provides a general formulation of ABC model choice using discrepancies, allowing different distance measures to be employed within a unified framework. We consider several statistical distance metrics suitable for this framework. All belong to the class of integral probability semimetrics, so the theoretical results of \cite{legramanti2023} apply for parameter inference. These discrepancies are used in the simulation study and real data application of Sections~\ref{sec:simu} and ~\ref{sec:real} to assess the performance of full data ABC methods for model selection. 

The maximum mean discrepancy (MMD) \citep{park2015} compares embedded empirical distributions in a reproducing kernel Hilbert space, using a kernel $g: \mathcal{Y} \times \mathcal{Y} \rightarrow \mathbb{R}$, 
$$MMD^2(\mu_{0,\bm{\theta}_0}, \mu_{k^*,\bm{\theta}_{k^*}}) = \mathbb{E}[g(y_1, y_2)] +  \mathbb{E}[g(z_1, z_2)] -2\mathbb{E}[g(y_1, z_1)].$$ 
The unbiased empirical estimate is:
\begin{align}
MMD^2(\hat{\mu}_{0,\bm{\theta}_0}, \mu_{k^*,\bm{\theta}_{k^*}}) &= \frac{1}{n(n-1)} \sum_{i=1}^n \sum_{j \neq i} g(y_i, y_j)  +\frac{1}{n(n-1)} \sum_{i=1}^n \sum_{j \neq i} g(z_i, z_j) + \nonumber \\
& - \frac{2}{n^2} \sum_{i=1}^n \sum_{j=1}^n g(y_i, z_j). \label{eq:MMD}
\end{align}
A common choice of kernel is the Gaussian kernel,  $g(y, z) = \exp\left(-\frac{||y-z||_2^2}{2\sigma}\right)$, which produces an MMD that attempts to match all moments between the two distributions. A similar approach using the energy distance, a specific case of MMD where the Euclidean distance is used as the kernel function, was explored by \cite{nguyen2020}. 

\cite{bernton2019} proposed using the Wasserstein distance in ABC. For univariate data and assuming that $P_{M_0,\bm{\theta}_0}^{(n)}$ and $P_{M_{k^*},\bm{\theta}_{k^*}}^{(n)}$ have finite $p$-th moment with $p \geq 1$, the $p$-Wasserstein distance between these measures is given by 
$$\mathcal{W}_p(\mu_{0,\bm{\theta}_0}, \mu_{k^*,\bm{\theta}_{k^*}}) = \left(\int_0^1 |F^{-1}_{\mu_{0,\bm{\theta}_0}}(\lambda) -F^{-1}_{\mu_{k^*,\bm{\theta}_{k^*}}}(\lambda) |^p d\lambda \right)^{1 / p},$$ 
where $F_{\mu_{0,\bm{\theta}_0}}(\cdot)$ and $F_{\mu_{k^*,\bm{\theta}_{k^*}}}(\cdot)$ are the cumulative distribution functions of $P_{M_0,\bm{\theta}_0}^{(n)}$ and $P_{M_{k^*},\bm{\theta}_{k^*}}^{(n)}$, respectively. The empirical $p$-Wasserstein distance, with $p=1$, is computed by comparing order statistics:
\begin{equation}
\mathcal{W}_1(\hat{\mu}_{0,\bm{\theta}_0}, \hat{\mu}_{k^*,\bm{\theta}_{k^*}}) = n^{-1}\sum_{i=1}^n |y_{(i)} - z_{(i)}|.
\label{eq:Wass}    
\end{equation}
Although theoretically appealing because it can lead to an ABC algorithm that approximates the true posterior distribution arbitrarily well as $\varepsilon \rightarrow 0$, its performance may degrade when data quantiles are insensitive to parameters changes \citep{drovandi2022}.

The Cramér–von Mises (CvM) distance \citep{frazier2020}, motivated by minimum distance estimation \citep{donoho1988}, which provides nice properties such as yielding robust and potentially efficient point estimators, measures the $L_2$ difference between the empirical CDF $\hat{F}_{\mu_{k^*,\bm{\theta}_{k^*}}}$ and the CDF of the theoretical distribution $F_{\mu_{0,\bm{\theta}_0}}$
\begin{align*}
\mathcal{C}^2(&\mu_{0,\bm{\theta}_0}, \hat{\mu}_{k^*,\bm{\theta}_{k^*}})  = \int_{\mathcal{Y}}\left[\hat{F}_{\mu_{k^*,\bm{\theta}_{k^*}}}(y) - F_{\mu_{0,\bm{\theta}_0}}(y)\right]^2dF_{\mu_{0,\bm{\theta}_0}}(y). 
\end{align*}
It is estimated by:
\begin{align}
    \mathcal{C}^2(\hat{\mu}_{0,\bm{\theta}_0}, \hat{\mu}_{k^*,\bm{\theta}_{k^*}}) = \frac{U}{2n^2} - \frac{4n^2 - 1}{12n},
\label{eq:CvM}
\end{align}
where $U$ involves the ranks of $\bm{y}$ and $\bm{z}$ in their pooled sample.

\subsection{Theoretical justification}

Consider the version of Algorithm \ref{alg:full-data_abc-mc} using the Wasserstein distance and a sequence $\varepsilon_n \rightarrow 0$ of threshold varying with the sample size $n$, and define this algorithm ABC-Wass. Then, the ABC posterior probability of model $M_{k^*} \in \{M_1, \ldots, M_K\}=\mathcal{M}$ is 
\begin{equation}
    \pi_{\epsilon_n}(M_{k^*} \mid \bm{y}) \propto \pi(M_{k^*}) \int_{\Theta_{k^*}} \pi(\bm{\theta}_{k^*} \mid M_{k^*}) \, \mathbb{P}\left\{ \mathcal{W}_p(\hat{\mu}_{0,\boldsymbol{\theta}_0}, \hat{\mu}_{k^*,\boldsymbol{\theta}_{k^*}} ) \leq \varepsilon_n \right\} \, d\bm{\theta}_{k^*},
\end{equation}
where the left hand side is normalized over models.

The expression $\mathbb{P}(\cdot)$ refers to the random probability over simulations that a sample $\bm{\theta}_{k^*}$ under model $M_{k^*}$ yields a Wasserstein distance smaller or equal to a threshold $\varepsilon_n$. This is the noisy version of ABC considered in the framework of \cite{bernton2019}.

Under regularity conditions, ABC-Wass yields consistent parameter estimation \citep{bernton2019}. This property extends to model selection via Theorem \ref{th:consistency}, when the candidate models are well-separated in the Wasserstein sense, i.e. they generate distinguishable empirical distributions, and the Wasserstein distance between empirical samples is accurate enough. To guarantee this, the acceptance threshold must be taken as a sequence $\varepsilon_n \rightarrow 0$ as $n \rightarrow \infty$, since with growing sample size the empirical distributions concentrate around their population laws, and a fixed tolerance $\varepsilon$ would allow incorrect models to pass the ABC criterion whenever their Wasserstein distance from the true model is below $\varepsilon$.

Moreover, we restrict attention to the case of i.i.d. observations. This choice aligns with Assumption (A4) of Theorem \ref{th:consistency} on the convergence of Wasserstein distances, which holds under mild conditions, e.g. existence of finite $p$-th moments, in the i.i.d. setting. \citet{legramanti2023} extend certain convergence results to non-i.i.d. contexts, including time series models, but we leave such extensions for future work. 

\medskip

\begin{theorem}[Model Selection Consistency of ABC-Wass]
\label{th:consistency}
Assume:
\begin{enumerate}
    \item[(A1)] (Identifiability): For all $M_{k^*} \neq M_0$, there exists $\delta_{k^*} > 0$ such that 
    $$\inf_{\bm{\theta}_{k^*} \in \Theta_{k^*}}\mathcal{W}_p(P^{(n)}_{M_0,\bm{\theta}_0}, P^{(n)}_{M_{k^*},\bm{\theta}_{k^*}}) > \delta_{k^*}.$$
    \item[(A2)] (Prior Positivity): $\pi(M_0) > 0$ and $\pi_{M_0}(\bm{\theta}_0) > 0$.
	\item[(A3)] (Tail Behavior): For each $M_{k^*} \neq M_0$, let $\mathcal{B}_{\delta_{k^*}} = \{\bm{\theta}_{k^*} \in \Theta_{k^*}: \mathcal{W}_p(P_{M_0,\bm{\theta}_0}, P_{M_{k^*},\bm{\theta}_{k^*}}) \leq 2\delta_{k^*}/3 \}$, where $\delta_{k^*}$ is as in Assumption (A1). Then, for any sequence $\varepsilon_n \to 0$ with $\varepsilon_n < \delta_{k^*}/2$ for large $n$
    \[
    \int_{\Theta_{k^*} \setminus \mathcal{B}_{\delta_{k^*}}} \pi_{M_{k^*}}(\bm{\theta}_{k^*}) \cdot \mathbb{P}(\mathcal{W}_p(\hat{\mu}_{0,\bm{\theta}_0}, \hat{\mu}_{k^*,\bm{\theta}_{k^*}}) \leq \varepsilon_n) d\bm{\theta}_{k^*} \to 0 \quad \text{as } n \to \infty.
    \]
    \item[(A4)] (Empirical Convergence): $\mathcal{W}_p(\hat{\mu}_{0,\bm{\theta}_0}, P^{(n)}_{M_0,\bm{\theta}_0}) \to 0$ and $\mathcal{W}_p(\hat{\mu}_{k^*,\bm{\theta}_{k^*}}, P^{(n)}_{M_{k^*},\bm{\theta}_{k^*}}) \to 0$ almost surely as $n \to \infty$, and $\Delta_n := \sup_{M_{k^*}, \bm{\theta}_{k^*}} \mathbb{E}[\mathcal{W}_p(\hat{\mu}_{k^*,\bm{\theta}_{k^*}}, P^{(n)}_{M_k^*,\bm{\theta}_{k^*}})]$ satisfies $\Delta_n = o(\varepsilon_n)$. 
\end{enumerate}
Then, for any sequence $\varepsilon_n \to 0$ such that $\varepsilon_n < \min_{k^*} \delta_{k^*} / 2$,
\[
\pi_{\varepsilon_n}(M_0 \mid \bm{y}) \overset{p}{\to} 1 \quad \text{as } n \to \infty.
\]
\end{theorem}

The proof of Theorem \ref{th:consistency} is provided in Appendix B of the Supplementary Material. Theorem \ref{th:consistency} states that, under suitable conditions, the ABC posterior over models assigns probability one to the true model $M_0$ as the sample size increases. Assumption (A1) ensures that for any incorrect model $M_{k^*} \neq M_0$ and any parameter $\bm{\theta}_{k^*} \in \Theta_{k^*}$, the Wasserstein distance between the simulated data distribution $P^{(n)}_{M_{k^*},\bm{\theta}_{k^*}}$ and the true distribution $P^{(n)}_{M_0,\bm{\theta}_0}$ is bounded below by some fixed $\delta_{k^*}>0$. This guarantees that no parameter in a wrong model can generate data arbitrarily close to the true data; without this, a wrong model might mimic the observed data. Assumption (A2) requires that both the model prior and the parameter prior assign positive mass to a nighborhoods of the true model and true parameter. Assumption (A3) ensures that parameter values far (in Wasserstein distance) to $P^{(n)}_{M_0,\bm{\theta}_0}$ cannot produce empirical measures within the small threshold with non-negligible prior mass. In other words, the prior must decay sufficiently fast in regions where the Wasserstein distance between the simulated and true distribution is large. Finally, Assumption (A4) holds for i.i.d. observations and in cases with a finite $p$-th moment.

\begin{theorem}[Robustness to Model Misspecification]
\label{thm:misspec}
Let the true data distribution $P^{(n)}_{M_0,\bm{\theta}_0}$ be misspecified with respect to the candidate set
$$\{P^{(n)}_{M_k,\bm{\theta}_k}: M_k \in \mathcal{M}, \bm{\theta}_k \in \Theta_k\}.$$ 
Suppose Assumptions (A1)-(A4) hold, and let the ABC tolerance decrease to zero, i.e. $\varepsilon_n \to 0$, satisfying $\varepsilon_n n^{\beta} \to \infty$ where $\beta > 0$ is such that $\mathcal{W}_p(\hat{\mu}_{0},P^{(n)}_{M_0,\bm{\theta}_0)} = O_{\mathbb{P}}(n^{-\beta})$. Then the ABC-Wass posterior asymptotically concentrates on the model–parameter pair 
\[
(M^\dagger, \bm{\theta}^\dagger) = \arg\min_{M_{k} \in \mathcal{M}, \bm{\theta}_{k} \in \Theta_{k}} \mathcal{W}_p(P^{(n)}_{M_0,\bm{\theta}_0},P^{(n)}_{M_{k},\bm{\theta}_{k}}).
\]
\end{theorem}

The proof of Theorem \ref{thm:misspec} is available in Appendix C of the Supplementary Material. This result shows that, under model misspecification, ABC-Wass asymptotically selects the model whose distribution is closest to the true data distribution in the Wasserstein metric.

We evaluate full data ABC approaches under the i.i.d. assumption in Sections \ref{sub:expo}–\ref{sub:gandk} and extend the analysis to dependent data in Sections \ref{sub:toad}–\ref{sec:real}. Although Wasserstein, MMD, and CvM distances are traditionally defined for i.i.d. samples, they can be adapted to dependence, for example via time-aware kernels or block structures. \citet{legramanti2023} establishes that, under suitable regularity conditions, the ABC posterior based on such distances retains uniform convergence properties in the temporally dependent setting.

\section{Simulation Study}
\label{sec:simu}

The goal of this section is to assess ABC’s empirical performance with various distances and to compare it with other methods.

This section applies ABC with statistical distances (Algorithm \ref{alg:full-data_abc-mc}), specifically MMD \eqref{eq:MMD}, Wasserstein \eqref{eq:Wass}, and CvM \eqref{eq:CvM}, to several simulated model selection problems, comparing performance against summary-based ABC (Algorithm 3, Appendix A), in which problem-specific summary statistics are used. The full data approaches are referred to as ABC-MMD, ABC-Wass, and ABC-CvM, while ABC-Stat denotes the summary-based method.

Each experiment is repeated 100 times for sample sizes $n=100$ and $n=1000$ in Section \ref{sub:normal}, \ref{sub:expo} and \ref{sub:gandk}, and it is repeated 100 times for a sample sizes replicating the true data example of Section \ref{sec:real} in Section \ref{sub:toad}. For each observed dataset, $10^6$ simulations are generated to approximate posterior model probabilities and model-specific parameter posteriors for all examples, except the one described in Section \ref{sub:toad}. Model priors are taken as uniform, $\pi(M=M_k)=1/K$ for $k=1,\dots,K$, and parameter priors are defined individually for each example. The distance threshold $\varepsilon$ is not fixed a priori; instead, following \cite{biau2013}, it is chosen as the $q$-th percentile of the simulated distances, retaining only the closest $q\%$ of simulations. Three values of $q$ were recorded: $q=\{0.1,0.5,0.01\}$. 

Across all examples, performance is assessed using: (i) the estimated probability of selecting the correct model; (ii) the average posterior means of the parameters and their mean squared errors (MSEs); (iii) classification accuracy summarised through confusion matrices (shown in the Appendixes); and (iv) boxplots of the posterior probability of selecting the correct model across repetitions (shown in the Appendixes).

To further assess ABC's effectiveness in model choice and parameter inference, we compare it with a deep neural network (NN) trained on the same models and tested on the same 100 datasets. The network performs multi-task learning: classification (model selection) with categorical cross-entropy loss and regression (parameter estimation) using MSE loss. Loss weights are 1.0 (classification) and 0.5 (regression), prioritizing model identification. The architecture comprises five fully connected layers.

The neural network is trained on $10^6$ simulated datasets, with 80\% used for training and 20\% for testing. To ensure a controlled and meaningful comparison, training data are generated under the same candidate models considered by the ABC procedures. In contrast to ABC, which simulates parameters from prior distributions and selects samples closest to the observed data according to a discrepancy measure, the NN learns a discriminative mapping from data directly to models and parameters. As such, the NN represents a fundamentally different inferential paradigm, relying on supervised learning rather than simulation-based posterior approximation.

In our experiments, the NN is trained using datasets generated from fixed parameter values for each model, assuming that the true model is included in the simulation design. In practical likelihood-free settings, parameters are typically uncertain and drawn from prior distributions, which induces additional within-model variability. We therefore also conducted a comparison in which the NN was trained using parameters drawn from the same priors as those used by the ABC methods.

When training the neural network using parameters drawn from the prior distributions, we observe different effects depending on the nature of the inferential problem. In the hypothesis testing examples involving a point null hypothesis (Section \ref{sub:normal} and \ref{sub:gandk}), NN performance improves when parameters are sampled from the prior, reflecting the fact that prior-based training better represents the marginal evidence under the alternative hypothesis. In contrast, for the more general model selection examples considered in this paper, where each model corresponds to a distinct data-generating mechanism rather than a nested alternative, training the NN under parameter priors introduces substantial within-model variability and leads to reduced classification accuracy relative to training with fixed parameter values.

These findings highlight an important conceptual distinction between ABC and NN-based approaches. ABC explicitly targets the marginal evidence by integrating over parameter uncertainty through simulation from the prior and comparison of generated datasets to the observed data. NNs, by contrast, learn a discriminative approximation to the model posterior, and their performance depends sensitively on how parameter uncertainty is represented in the training data. As a result, while NN-based methods can perform well in hypothesis testing settings where prior integration sharpens evidence against a point null hypothesis, their behavior in general model selection problems is more problem-dependent.

Finally, we include two additional methods for comparison:
(1) the semi-automatic summary selection method (ABC-SA) of \cite{prangle2013}, implemented in the \texttt{abctools} R package \citep{nunes2015abctools}; and
(2) the classification-based ABC using quadratic discriminant analysis (ABC-QDA) from \cite{gutmann2018likelihood}, originally developed for parameter inference rather than model selection. We select QDA as the classification algorithm because \cite{gutmann2018likelihood} demonstrated its robustness across multiple examples. We include this method as an additional classification-based benchmark, although it was originally developed for parameter inference rather than model selection.
  
\subsection{Normal Mean Hypothesis Test}
\label{sub:normal}

We assess ABC using distance metrics for model choice in a simple normal-data setting. Let $\bm{y} = (y_1, \dots, y_n)^T$ be i.i.d. with $y_i \sim N(\theta, \sigma^2)$, assuming $\sigma^2 = 1$. The competing models correspond to $H_0: \theta = \tilde{\theta}$ versus $H_1: \theta \neq \tilde{\theta}$.

Synthetic datasets are generated from the true model $\mathcal{N}(\theta_0, 1)$ with $\theta_0 \in \{0.0, 0.1, 0.2, \\ 0.3, 0.4, 0.5\}$, allowing evaluation of method performance as the true mean departs from $\tilde{\theta}=0$. For each value of $\theta_0$, 100 datasets are simulated and analyzed using the seven methods described above for model selection and parameter inference. Under $H_1$, the prior is $\theta \sim N(\tilde{\theta}, 100)$, providing a weakly informative prior; under $H_0$, $\theta$ is fixed at $0$.

For this simple normal example with known variance, the Bayes factor $B_{01}$ can be computed analytically \citep[Chapter 5]{robert2007bayesian}. While we do not use it in our evaluation, this example provides a context where the ABC procedure can, in principle, be compared to exact Bayesian inference. For ABC-Stat, we use the sufficient statistic $\eta(\bm{y}) = \bar{y}$ and the Euclidean distance.

We first assess each method’s ability to correctly identify the data-generating model. Performance is summarized using the estimated probability of selecting $H_0$ across different values of $\theta_0$ (Table \ref{ex_normal_table}), and confusion matrices and boxplots of posterior probabilities (Appendix D of the Supplementary Material). ABC methods based on distance metrics correctly identify $H_0$ in all cases, indicating good control of false positives. The transition is gradual: for small deviations from zero (e.g., $\theta_0 = 0.1$), the classifier shows limited power, but confidence in selecting $H_1$ grows as the true mean increases. When $\theta_0 \ge 0.4$, $H_1$ is selected most of the time, reaching around 95\% selection accuracy at $\theta_0 = 0.5$. ABC-SA follows a similar pattern but tends to select $H_0$ more often than the full data ABC approaches as the true mean increases. ABC-QDA selects $H_0$ in nearly all cases. NN performance depends on how parameter uncertainty under the alternative hypothesis is represented during training. When trained using fixed parameter values, NN detects departures from the null for moderate values of $\theta_0$ (from about $\theta_0=0.2$), but shows high variability and poor calibration near the decision boundary. Training under parameters drawn from the prior improves stability and yields classification accuracy comparable to the ABC methods.

We further evaluate posterior inference quality for $\theta$ by computing the MSE between the posterior mean and the true value $\theta_0$ across repetitions (Table \ref{ex_normal_table}). For each dataset, the posterior mean is computed from the model selected as most probable by each method. Results show that MSE is generally low, with the smallest values observed when $\theta_0 = 0$ and $\theta_0 = 0.5$, consistent with improved classification accuracy for these values. When the true mean is small but non-zero, posterior distributions for all methods retain substantial mass near zero, reflecting limited information in the data to distinguish between models. ABC-SA and ABC-QDA exhibit larger MSE than the distance-based ABC methods. NN achieves low MSE across all cases, particularly when trained under parameter priors, although this reflects the simplicity of the setting and the availability of training data that closely matches the true generative mechanism.

All methods perform similarly when the sample size is $n = 1{,}000$, showing minimal differences in both model selection and parameter estimation. ABC methods also yield comparable results across smaller tolerance thresholds; therefore, these additional results are omitted.

These results demonstrate that ABC with distance metrics provides reliable performance for both model selection and parameter estimation, comparable to ABC-Stat using sufficient summary statistics. The comparison with other ABC algorithms in this simple example highlights the limitations of those methods and illustrates the advantages of full data ABC approaches for model selection problems. The NN comparison illustrates that discriminative approaches can perform well in hypothesis testing problems when parameter uncertainty is properly incorporated during training. However, as shown in the following sections, this favorable behavior does not extend uniformly to more complex model selection problems considered in this paper, where ABC methods exhibit more stable and robust performance.

\begin{table}[]
\footnotesize
\begin{tabular}{c|c|cccc|cc|cc}
      &             & \multicolumn{4}{c|}{\textbf{ABC}}                          & \multicolumn{2}{c|}{\textbf{NN}} & \multicolumn{2}{c}{\textbf{ABC}} \\
      &             & \textbf{CvM} & \textbf{MMD} & \textbf{Wass} & \textbf{Stat} & \textbf{fixed} & \textbf{prior}& \textbf{SA}                 &  \textbf{QDA}                 \\
      \hline
$\theta_0=0.0$      & $\Pr(\theta=0)$        & 0.975                & 0.976                & 0.975                 &            0.973    &  0.599     & 0.979 & 0.976               &  0.981         \\
	& $\hat{\theta}$ & 0.000                & 0.000                & 0.000                 & 0.002            & 0.042    & 0.010   & 0.000               & 0.000                \\
      & MSE         & 0.000                & 0.000                & 0.000                 & 0.001            & 0.002 & 0.012    & 0.000               & 0.000                \\
      \hline
$\theta_0=0.1$      & $\Pr(\theta=0)$        & 0.931                & 0.933                & 0.929                 &            0.412    &  0.925  & 0.926   & 0.948               &  0.974         \\

& $\hat{\theta}$ & 0.007           & 0.006           & 0.007            & 0.010            & 0.052       & 0.013 & 0.000               & 0.000                \\
      & MSE         & 0.011           & 0.011           & 0.011            & 0.011            & 0.002     & 0.008 & 0.010          & 0.010           \\
\hline
$\theta_0=0.2$      & $\Pr(\theta=0)$        & 0.788                & 0.796                & 0.789                 &            0.760    &  0.261   & 0.747  & 0.847               &  0.955         \\
& $\hat{\theta}$ & 0.046           & 0.042           & 0.042           & 0.048            & 0.155     & 0.027 & 0.010          & 0.000                \\
      & MSE         & 0.037           & 0.037           & 0.038            & 0.038            & 0.013     & 0.031  & 0.037          & 0.040           \\
\hline
$\theta_0=0.3$      & $\Pr(\theta=0)$        & 0.522                & 0.543                & 0.519                 &            0.459    &  0.084  & 0.462   & 0.671               &  0.934         \\
& $\hat{\theta}$ & 0.147           & 0.144           & 0.149            & 0.168            & 0.273     & 0.049 & 0.141          & 0.000                \\
      & MSE         & 0.062           & 0.061           & 0.060            & 0.056            & 0.004 & 0.063 & 0.096          & 0.090           \\
\hline
$\theta_0=0.4$      & $\Pr(\theta=0)$        & 0.210                & 0.236                & 0.203                 &            0.157    &  0.030   & 0.196  & 0.450               &  0.902         \\
& $\hat{\theta}$ & 0.344           & 0.335           & 0.345            & 0.364            & 0.397      & 0.059 & 0.285          & 0.000                \\
      & MSE         & 0.039           & 0.041           & 0.038            & 0.034            & 0.003      & 0.117 & 0.125          & 0.160           \\
\hline
$\theta_0=0.5$      & $\Pr(\theta=0)$        & 0.056                & 0.067                & 0.050                 &            0.031    &  0.014   & 0.059  & 0.263               &  0.870         \\
& $\hat{\theta}$ & 0.494           & 0.489           & 0.497            & 0.506            & 0.484          & 0.078 & 0.359          & 0.000                \\
      & MSE         & 0.022           & 0.025           & 0.020            & 0.016            & 0.003     & 0.178  & 0.125          & 0.250          
\end{tabular}
\caption{Summary of results for the normal model across repetitions. For each true mean, the table reports the average probability of selecting hypothesis $H_0$, the average posterior mean, and the corresponding MSE over all repetitions. For all ABC methods, $q\%=0.1\%$ is used. Similar results are obtained with lower percentiles.}
\label{ex_normal_table}
\end{table}

\subsection{Exponential Family Model Selection}
\label{sub:expo}

In this example, we adopt the same model setup as in \cite{marin2016} and consider three models from the exponential family for which the marginal likelihoods are available in closed form, allowing for analytical computation of the Bayes factors and posterior model probabilities. In our study, these quantities are not used, but we retain the model specification to facilitate comparison with their results.

Model $M_1$ assumes the data follow an exponential distribution with rate parameter $\theta$,
$y_i \stackrel{\mathrm{i.i.d.}}{\sim} \mathcal{E}xp(\theta)$, $i=1,\dots,n$, with prior $\theta \sim \mathcal{E}xp(1)$.
Model $M_2$ specifies a log-normal distribution with location parameter $\theta$ and dispersion fixed at 1,
$y_i \stackrel{\mathrm{i.i.d.}}{\sim} \mathcal{LN}(\theta,1)$, with prior $\theta \sim N(0,1)$.
Model $M_3$ assumes a gamma distribution with shape fixed at 2 and unknown rate parameter $\theta$,
$y_i \stackrel{\mathrm{i.i.d.}}{\sim} Ga(2,\theta)$, with prior $\theta \sim \mathcal{E}xp(1)$.

With these distributions belonging to the exponential family, the following choice of summary statistics for ABC-Stat is sufficient for each model and across models in the sense of \cite{marin2013}. This means that, for the model choice problem, no information is lost by replacing the full dataset with these summaries:
$$
\bm{\eta}(\bm{y}) = \left(\sum_{i=1}^n y_i, \sum_{i=1}^n \log y_i, \sum_{i=1}^n \log^2 y_i\right).
$$
To assess the consistency of each ABC method, we fix $\theta$ when generating the observed data so that the true expected value under each model equals 2. This corresponds to $\theta = 1/2$ for $M_1$, $\theta = \log 2 - 1/2$ for $M_2$, and $\theta = 1$ for $M_3$. The resulting densities are quite similar, which makes model selection more challenging.

Given the positive skewness observed in the densities, we apply log-transformations to both observed and simulated data for ABC-MMD and ABC-Wass. This follows the approach in \cite{drovandi2022} for parameter estimation in skewed datasets. Since the CvM distance is invariant to monotonic transformations, no transformation is applied for ABC-CvM.

Results are summarised in Table \ref{ex_expo_table} and Appendix E of the Supplementary Material. When the true model is $M_1$ or $M_2$, achieving high posterior probability for the correct model is challenging in some cases, whereas $M_3$ is generally identified with higher confidence by all methods. Applying a log transformation to the data improves the performance of ABC-Wass and ABC-MMD, enabling them to match the accuracy of ABC-Stat (with a posterior probability of the correct model close to one). ABC-Stat consistently achieves near-exact results. Without log transformations, MMD and Wasserstein performed similarly to CvM, with noticeably higher classification error. This is a significant observation, showing that distance-based ABC can approach optimal classification even without explicitly sufficient summaries. For $n=1000$, all ABC-based methods correctly identified the true model in every case, and we therefore omit the results.

ABC-SA performs worse than ABC-Stat, likely due to its reliance on general-purpose summary statistics rather than sufficient problem-specific sufficient summaries, available in this specific case.

ABC-QDA shows variable performance across ABC methods, with correct model selection probabilities below 90\% for simulations from $M_2$ and $M_3$. Its reliance on QDA to separate observed and simulated data, followed by selection of simulations with low classification accuracy, appears better suited to parameter inference than model choice. Under misspecification, incorrect models can closely mimic the true model \citep{berk1966limiting}, and a possible explanation of ABC-QDA behavior is that the estimation process concentrates to parameter values that make the model closer to the true one, leading to lower classification accuracy and higher MSE for parameter estimates.

NN is the method that shows the lowest accuracy of model selection: it correctly identifies the exponential model in only 55\% of cases, the lognormal in 65\%, and the gamma in around 97\%. The lognormal model is frequently misclassified as exponential, suggesting a structural limitation in distinguishing these distributions. Training the NN using parameters drawn from the prior does not lead to the same improvement observed in the normal hypothesis testing example. Unlike Section 4.1, the present setting involves non-nested models with highly overlapping marginal distributions, and sampling parameters from the prior substantially increases within-model variability without sharpening the decision boundary between models. As a result, prior-based training makes the discriminative task more challenging and leads to reduced classification accuracy. As a remark, when training the NN using parameters drawn from the prior, we apply a component-wise 
$\text{arcsinh}(\cdot)$ transformation to the input data to mitigate the effect of extreme observations arising from prior draws with large parameter values. This normalization substantially stabilizes training but does not fully overcome the intrinsic difficulty of discriminating between highly overlapping models in this setting.

Parameter estimation results (Table \ref{ex_expo_table}) mirror these findings. For distance-based ABC methods, MSEs are consistently low when the correct model is selected, comparable to ABC-Stat. NN produces slightly more biased estimates than the ABC methods, in particular for datasets generated from the exponential and the lognormal models, reflecting its sensitivity to misclassification errors. ABC-SA yields slightly higher estimation errors. Finally, ABC-QDA has the largest MSE values across all experiments and ABC methods.

\begin{table}[h]
\hspace*{-1.5cm} 
\centering
\footnotesize
\begin{tabular}{c|c|cccccc|cc|cc}
      &             & \multicolumn{6}{c|}{\textbf{ABC}}                          & \multicolumn{2}{c|}{\textbf{NN}} & \multicolumn{2}{c}{\textbf{ABC}} \\
      &             & \textbf{CvM} & \textbf{MMD} & \textbf{MMD} & \textbf{Wass} & \textbf{Wass}& \textbf{Stat} & \textbf{fixed} & \textbf{prior} & \textbf{SA}                 &  \textbf{QDA}                 \\
      &             & \multicolumn{2}{c}{}                          & \multicolumn{1}{c}{\textbf{(log)}} & \multicolumn{1}{c}{} & \multicolumn{1}{c}{\textbf{(log)}} & \multicolumn{1}{c|}{}  & \multicolumn{2}{c|}{} & \multicolumn{2}{c}{} \\
      \hline
$\mathcal{E}\text{xp}(0.5)$      & $\Pr(M=M_1)$        & 0.883                & 0.836                & 0.953 & 0.850     & 0.948    &            0.952    &  0.554   &   0.453 & 0.772          &    0.922       \\
	& $\hat{\theta}$ & 0.512                & 0.506                &  0.511 & 0.515   & 0.514              & 0.513            &  0.454      &  0.342 &   0.510           &    0.522 \\
      & MSE         & 0.003              & 0.003                & 0.003 & 0.003  & 0.003               & 0.003            & 0.021      &   0.028 &  0.003           &                0.004 \\
      \hline
$\mathcal{LN}(0.193,1)$      & $\Pr(M=M_2)$        & 0.896                & 0.829     & 0.952           & 0.882        & 0.956         &            0.954    & 0.653   & 0.549   &     0.787           &     0.875      \\

& $\hat{\theta}$ & 0.188           & 0.196          & 0.187 & 0.189    & 0.187        & 0.191            & 0.337      & 0.367 &   0.204            &                0.187 \\
      & MSE         & 0.009           & 0.008          & 0.008 & 0.008     & 0.009       & 0.010            & 0.034      &   0.038 &  0.013      &    0.140       \\
\hline
$\mathcal{G}\text{a}(2,1)$      & $\Pr(M=M_3)$        & 0.952               & 0.966    & 0.971            & 0.984      & 0.987           &            0.987    & 0.967  & 0.934    &   0.955             &   0.819     \\
& $\hat{\theta}$ & 1.000           & 0.995          & 0.998 & 1.004     & 1.003      & 1.002            &  0.957  & 0.946   &  1.017        &    1.000            \\
      & MSE         & 0.004          & 0.004          & 0.004 & 0.004   & 0.004         & 0.004            &  0.008  & 0.017   &  0.008         &    0.275       
\end{tabular}
\caption{Results for the exponential family models across repetitions. For each model, we report the average posterior probability of selecting the correct model, the average estimate of the parameter $\theta$, and the MSE across repetitions. For all ABC methods, $q\%=0.1\%$ is used. Similar results are obtained with lower percentiles.}
\label{ex_expo_table}
\end{table}

\subsubsection{Computational Performance Comparison}

The proposed approaches exhibit varying computational costs. For both NN and ABC methods, we generated $10^6$ datasets, resulting in similar training times.

For $n=100$, the computation times for distance calculations between one observed dataset and one simulated dataset are approximately 0.0018 s for ABC-CvM, 0.0001 s for ABC-Wass, and 0.2685 s for ABC-MMD. For $n=1000$, the times are approximately 0.0040 s for ABC-CvM, 0.0013 s for ABC-Wass, and 0.2738 s for ABC-MMD. As a comparison, ABC-Stat takes about 0.0053 s for $n=100$ and 0.0582 s for $n=1000$, showing similar computational cost to ABC-CvM for the examples considered in this paper. Hence, ABC-MMD is the most computationally demanding, with running times more than 2,000 times slower than ABC-Wass and almost 150 times slower than ABC-CvM. This is because standard MMD computation requires pairwise kernel evaluations between all points in both datasets, resulting in $\mathcal{O}(n^2)$ complexity. In contrast, metrics like the Wasserstein distance can often be computed more efficiently, especially in one dimension. Moreover, MMD relies on kernel functions mapping data into a reproducing kernel Hilbert space (RKHS), adding further computational burden. For comparison, ABC-QDA \citep{gutmann2018likelihood} takes 0.0529 s to perform QDA on the observed and simulated datasets, making it more computationally demanding than ABC-CvM and ABC-Wass.

The computational efficiency of ABC-Stat depends heavily on the number and nature of summary statistics used to represent the data: simple, low-dimensional summaries make it comparable to ABC-CvM, while more complex or high-dimensional statistics can lead to a substantial increase in runtime. However, for the examples presented in this paper, where the chosen summaries are simple and low-dimensional, the computation performance of ABC-Stat is very fast.

For completeness, we also measured the computational cost of ABC-SA, which involves an initial summary-statistic selection step followed by ABC rejection. ABC-SA is not directly comparable with the other methods, because it was run on a different software and directly on the full set of input summary statistics. When implemented using the \texttt{selectsumm()} function from the \texttt{abctools} package in \texttt{R} with $10^6$ simulated datasets and all observations as input summary statistics, each full call requires 540.7 s for $n=100$ and 1620.9 s for $n=1000$. It should be noted, however, that this comparison is not directly equivalent to the other methods (which are Python-based), since the \texttt{R} implementation performs the entire selection and rejection step jointly across all simulations, and involves additional overhead from the interpreted environment. The runtime of ABC-SA scales approximately linearly with the number of simulations and with the dimension of the summary vector, since the selection step must evaluate all simulations jointly. Consequently, ABC-SA remains efficient for moderate-sized problems but may become slower for very high-dimensional summaries or larger simulation budgets.

Total runtime depends on available computational resources. In our example, the analyses were executed in parallel on a 24-core Intel Xeon Scalable `Cascade Lake' processor with 8.91 GB of DDR4 RAM. Processing 100 observed datasets, each compared against $10^6$ simulated datasets using six distance metrics, took approximately one hour. By contrast, the deep neural network with 5 layers required 256 s to train over 10 epochs and an additional 4 s for testing.

Subsequent examples show comparable computational times, so we omit further runtime comparisons.

\subsection{$g$-and-$k$ Distribution Model Selection}
\label{sub:gandk}

Being a standard benchmark in the ABC literature for problems with intractable likelihoods, we consider a model selection problem involving the $g$-and-$k$ distribution \citep{rayner2002}. This distribution is defined via its quantile function:
\begin{align*}
\mathcal{Q}(p; a, b, c, g, k) &= a + b\left(1 + c\frac{1-e^{-gz(p)}}{1+e^{-gz(p)}}\right) \cdot \left(1 + z(p)^2\right)^kz(p),
\end{align*}
where $z(p)$ denotes the quantile function of the standard normal distribution. The constant $c = 0.8$ ensures that $\mathcal{Q}$ defines a proper distribution. Parameters $a,b>0$ represent location and scale, $g$ controls skewness, and $k > -\tfrac{1}{2}$ controls kurtosis. The $g$-and-$k$ distribution is attractive for its flexibility, but the absence of a closed-form density complicates standard likelihood-based inference.

Following \citet{marin2013}, we fix $a = 0$, $b = 1$, and $c = 0.8$, and focus on model selection between two submodels: $M_1$: $g=0$ (no skewness) vs $M_2$: $g \neq 0$ with $g \sim U(0,4)$ (positive skewness). We treat $k$ as unknown under both models, with prior $k \sim U(-0.5,5)$. Thus, the task reduces to a hypothesis test for skewness in the presence of unknown kurtosis.

Observed datasets are simulated using $g=0$, $k=2$ under $M_1$ and $g=1$, $k=2$ under $M_2$, following \citet{marin2013}. These parameter settings yield models that produce similar data distributions, making classification deliberately challenging. Appendix F in the Supplementary Material illustrates the overlap between the distributions.

In the ABC-Stat method, following \cite{marin2013}, the 0.1 and 0.9 quantiles are used as summary statistics, and the $L_1$ distance is employed. This choice is theoretically justified to recover the correct model asymptotically under both $M_1$ and $M_2$. All reported results based on ABC methods use the 0.1\% acceptance threshold. Similar results are obtained with smaller thresholds. Since the observations can be negative, log-transformations are not feasible, and distances are therefore computed on the untransformed data.

Tables \ref{ex_quantile_table_n100} and \ref{ex_quantile_table_n1000}, and Appendix F in the Supplementary Material present results for this example, for sample sizes $n=100$ and $n=1000$. For $n=100$ and when $M_1$ is the true model, full data ABC methods yield an average estimated probability around 0.80, which increases to 0.90 when the sample size grows to $n=1000$. ABC-MMD performs slightly worse than the other methods, particularly for the larger sample size, where it still shows high variability (see boxplots Appendix F).

When the data are generated from $M_2$, the estimated probabilities for $n=100$ are lower and varying depending on the method, but they approach one as $n=1000$, with the exception of ABC-MMD. Overall, ABC-Wass outperforms the other methods, producing consistently higher estimates of the posterior probability of the correct model and converging to one more quickly as the sample size increases, indicating strong discriminative performance.
 
These results are supported by the confusion matrices in Appendix F of the Supplementary Material, which show that ABC methods based on distance metrics outperform the alternatives. Among them, ABC-Wass achieves the strongest performance, reaching near-perfect classification when $n=1000$.

The behavior of ABC-SA is apparently surprising. At $n=100$, its performance is comparable to full data ABC methods, but when $n=1000$ it systematically favors $M_1$ even when the data arise from $M_2$. However, to reduce dimensionality and memory usage in this case, only every fifth order statistic was retained at $n=1000$ as initial vector of statistics, which may have limited the discriminative power of the summaries. 

ABC-QDA performs poorly across all scenarios, with high variability in the estimated posterior probabilities. This behavior reflects its sensitivity to model misspecification, whereby parameters from an incorrect model can approximate the data closely enough to obscure clear classification boundaries; this behavior is also suggested by the estimated values of $g$ (Tables \ref{ex_quantile_table_n100} and \ref{ex_quantile_table_n1000}), which are associated with the largest MSEs.

Although some methods achieve strong classification performance, this does not always translate into accurate parameter recovery. For example, while ABC-Wass is highly effective for model selection, it is known to perform less well when estimating parameters under the $g$-and-$k$ model \citep{drovandi2022}, while ABC-CvM provides more accurate parameter inference. This pattern is confirmed in this example, where the MSE for estimating $g$ tends to be lower for ABC-CvM than ABC-Wass.

NN achieves the best identification performance for model $M_2$, but it struggles to correctly recognize data generated under $M_1$. The estimation of the $g$ parameter shows the lowest MSE when $M_2$ is true but the highest MSE under $M_1$. This pattern remains consistent across all sample sizes, indicating that increased data availability does not mitigate this bias. NN therefore performs well when the underlying asymmetry is identifiable, but it remains sensitive to overlapping model features. As in the normal mean example of Section \ref{sub:normal}, this hypothesis testing setting benefits from training the NN using parameters drawn from the prior: prior-based training improves classification performance, particularly when the data are generated under $M_1$, as it better represents the marginal evidence under the alternative hypothesis and reduces misclassification near the decision boundary. Nevertheless, in contrast to the normal example, the ABC methods based on statistical distances continue to exhibit stronger performance than the NN trained under the prior.

Across all scenarios, all methods show improved performance when estimating the $k$ parameter.

\begin{table}[h!]
\footnotesize
\begin{tabular}{c|c|cccc|cc|cc}
      &             & \multicolumn{4}{c|}{\textbf{ABC}}                          & \multicolumn{2}{c|}{\textbf{NN}} & \multicolumn{2}{c}{\textbf{ABC}} \\
      &             & \textbf{CvM} & \textbf{MMD} & \textbf{Wass} & \textbf{Stat} & \textbf{fixed} & \textbf{prior} & \textbf{SA}                 &  \textbf{QDA}                 \\
      \hline
$g=0.0$      & $\Pr(M=M_1)$        & 0.793                & 0.701                & 0.808                 &            0.788    &  0.286    &  0.672 &   0.774           &   0.582       \\
	& $\hat{g}$ & 0.164                & 0.342                & 0.291                 & 0.262            &   0.706     &  0.582 &  0.190             &  1.533               \\
    & MSE ($g$)         & 0.308 & 0.712 & 0.540 & 0.508            &   0.556    & 0.470 &   0.229             &   2.516               \\
  &  $\hat{k}$ & 2.018                & 2.001                & 1.995                 & 2.009            &    2.000   & 2.766 &  2.015              &  2.016               \\
    & MSE ($k$)         & 0.074 & 0.089 & 0.048 & 0.118           &   0.000  & 1.142  &      0.077          &     0.171             \\
    \hline
$g=1.0$      & $\Pr(M=M_2)$        & 0.735                & 0.656                & 0.832                 &            0.742    &  0.970   &    0.684   & 0.846          &   0.475       \\
	& $\hat{g}$ & 1.163                & 0.825                & 1.103                 & 1.205            &   0.952   & 1.076 &    0.678            &    1.738             \\
    & MSE ($g$)         & 0.890 & 1.035 & 1.127 & 1.086            &  0.004    & 0.387 &  0.460              &      0.734            \\
  &  $\hat{k}$ & 2.066                & 1.954                & 2.005                 & 2.040            &   2.000   & 2.623  &   3.759             &     2.131            \\
    & MSE ($k$)         & 0.089 &  0.104 & 0.051 & 0.133          &  0.000   & 0.980  &    3.112            &   0.265               \\    
\end{tabular}
\caption{Results for datasets of size $n=100$ generated from a $g$-and-$k$ distribution with unknown $g$ and $k$. For each model ($g=0$ and $g\neq0$), we report the average posterior probability of the correct model, the average posterior means of $g$ and $k$, and their mean squared errors (MSE). For all ABC methods, $q\%=0.1\%$ is used. Similar results are obtained with lower percentiles.} 
\label{ex_quantile_table_n100}
\end{table}

\begin{table}[h!]
\footnotesize
\begin{tabular}{c|c|cccc|cc|cc}
      &             & \multicolumn{4}{c|}{\textbf{ABC}}                          & \multicolumn{2}{c|}{\textbf{NN}} & \multicolumn{2}{c}{\textbf{ABC}} \\
      &             & \textbf{CvM} & \textbf{MMD} & \textbf{Wass} & \textbf{Stat} & \textbf{fixed} & \textbf{prior} & \textbf{SA}                 &  \textbf{QDA}                 \\      \hline
$g=0.0$      & $\Pr(M=M_1)$        & 0.930 & 0.844 & 0.942 & 0.914    &  0.135    &     0.884 & 0.895          &    0.614      \\
	& $\hat{g}$ & 0.000 & 0.025 & 0.000 & 0.000            &   0.605   & 0.038  &    0.000            &    1.308             \\
    & MSE ($g$)         & 0.000 & 0.020 & 0.000 & 0.000       &   0.384  & 0.021  &  0.000              &   1.933               \\
  &  $\hat{k}$ & 2.010 & 2.013 & 1.988 & 2.009             &  2.001      &     1.932 & 1.975         &     1.880            \\
    & MSE ($k$)         & 0.007 & 0.008 & 0.004 & 0.010             &   0.000   & 3.791 & 0.011              &       0.041           \\
    \hline
$g=1.0$      & $\Pr(M=M_2)$        & 0.991 & 0.775 & 0.999 & 0.981     &  0.999    &   0.941 & 0.152           &    0.386      \\
	& $\hat{g}$ & 1.012 & 0.955 & 1.2301 & 1.034            &   1.010     &   1.069 &   0.040         &     1.518            \\
    & MSE ($g$)         & 0.044 & 0.210 & 0.166 & 0.077 &  0.006     &      0.231        & 0.946 &    0.645              \\
  &  $\hat{k}$ & 2.027 & 2.018 & 1.988 & 2.008             &  2.004      &   2.058  & 1.833          &        2.010         \\
    & MSE ($k$)         & 0.010 & 0.013 & 0.008 & 0.014            &   0.000    &       4.292       & 0.052 &    0.046              \\    
\end{tabular}
\caption{Results for datasets of size $n=1000$ generated from a $g$-and-$k$ distribution with unknown $g$ and $k$. For each model ($g=0$ and $g\neq0$), we report the average posterior probability of the correct model, the average posterior means of $g$ and $k$, and their mean squared errors (MSE). For all ABC methods, $q\%=0.1\%$ is used. Similar results are obtained with lower percentiles.} 
\label{ex_quantile_table_n1000}
\end{table}

\subsection{Toad Movement Model Selection}
\label{sub:toad}

Amphibian movement data often display repeated site reuse interspersed with occasional long displacements, a behavior that motivates multiscale random walk (MRW) models \citep{gautestad2005}, which incorporates a power-law step distribution. Unlike memory-free random walks such as Lévy flights \citep{viswanathan1999, humphries2010, sims2014}, the MRW accounts for return steps and models the development of a cognitive map, leading to non-random spatial reuse. Although power-law models are flexible, a key challenge remains in effectively capturing return behavior, which is crucial for understanding amphibian movement.

Following \cite{marchand2017}, the random return step for toads occurs at the end of their nighttime foraging journey, when they perform a displacement $S_n$ from their last refuge location $Y_n$. The toad either moves to a new location, $Y_{n+1} = Y_{n} + S_n$, or returns to a previous daytime refuge site  ($Y_i$ for some $i=1, \ldots, n-1$). Under the MRW, the displacement follows a symmetric, zero-centered stable distribution, $S_n \sim S(\alpha, \gamma)$, defined via its characteristic function
$$\varphi(t; \alpha, \gamma) = e^{-|\gamma t|^\alpha},$$
where $0 \leq \alpha \leq 2$ is a stability parameter and $\gamma > 0$ is a scale parameter. Decreasing values of $\alpha$ lead to increasingly leptokurtic distributions, allowing rare long-distance events to be effectively modeled, which is ideal for capturing toad movements. Notably, the probability density function of this family is intractable, as it is given by the inverse Fourier transform of the characteristic function.

Despite the intractable density, the stable distribution can be sampled using the CMS algorithm \citep{chambers1976}, enabling posterior inference via ABC. We consider three candidate models for toad return behavior. For each model, the toad starts at $Y_1=0$, and the step sizes $S_n \sim S(\alpha,\gamma)$ are i.i.d. for $n=1, \ldots, d-1$. Individual toads are assumed to behave independently.

Model $M_1$ is the random return model, which assumes a constant probability of return, $p_0$, for the toad to any previously visited site. The return site is chosen randomly from all prior sites. Locations that have been visited multiple times naturally have higher chances of being revisited, allowing home range patterns to emerge. Under Model $M_1$, the random refuge location at the beginning of day $(n + 1) \in \{2, ..., n_d\}$ is given by
$$Y_{n + 1} =
\begin{cases}
Y_{n} + S_{n} &\text{with prob. } \; 1 - p_0, \\
Y_{i} &\text{with prob. } \; p_0 / n \;\; \forall \;i = 1, ..., n.
\end{cases}$$

Model $M_2$ also uses a constant probability of return $p_0$ but incorporates distance by assuming that, when the toad returns, it always chooses the nearest refuge site. Formally:
$$Y_{n + 1} =
\begin{cases}
Y_{n} + S_{n} &\text{with prob. } \; 1 - p_0, \\
\min\limits_{Y \in \{Y_1, ..., Y_n\}} |Y_{n+1} - Y| &\text{with prob. } \; p_0.
\end{cases}$$

Model $M_3$ is a distance-based return model, where the probability of returning to a given site $i$ decays exponentially with the distance $d_i$ from that refuge: $p_{ret(i)} = p_0 e^{-d_i/d_0}$, with $d_0$ controlling the strenght of the distance effect. Let $A_{n}$ denote the number of unique previous refuge sites at day $n$, $R_1,\ldots, R_{A_n}$, such that $R_i \neq R_j \; \forall i \neq j$ and each $R_i  = Y_j$ for some $j \in \{1,\ldots,n-1\}$. Multiple visits to the same location do not create new refuge sites. Assuming independence between previous sites, the movement model is  
$$Y_{n + 1} =
\begin{cases}
Y_{n} + S_{n} &\text{with prob. } \; \prod_{j = 1}^{A_{n+1}} (1 - p_{ret(j)}), \\
R_i &\text{with prob. } \; p_i \;\; \forall \; i = 1, ..., A_{n+1},
\end{cases}$$
where
$$
p_i = \frac{p_{ret(i)}}{\sum_{j=1}^{A_{n+1}} p_{ret(j)}} \;\left(1 - \prod_{j = 1}^{A_{n+1}} \left(1 - p_{ret(j)}\right)\right).
$$
If $Y_{n+1} = Y_n + S_n$, a new refuge site is created and $A_{n+2} = A_{n+1} + 1$; otherwise $A_{n+2} = A_{n+1}$.

For the ABC model selection procedure, given the higher computational burden of the simulation, we generated only $10^5$ simulated datasets, assuming a uniform prior over the three models. Parameter priors follow \cite{marchand2017}: $\alpha \sim U(1, 2)$, $\gamma \sim U(10, 100)$, $p_0 \sim U(0, 1)$, and $d_0 \sim U(20, 2000)$. 

To reduce the dimensionality of the observed data, we summarise it as follows, see \cite{drovandi2022}. For a set of time lags $\ell=1, 2, 4, 8$, we compute  displacements $\bm{y}_{\ell} = \{|\bm{y}_{i + \ell, j} - \bm{y}_{i, j}| \; | \; 1 \leq i \leq n_d - \ell, 1 \leq j \leq n_t\}$. Returns are defined as displacements smaller than 10 meters, i.e. $\big{|}\{x \in \bm{y}_\ell \; | \; x \leq 10 \}\big{|}$, and non-returns are displacements greater than or equal to 10 meters. For each lag, we extract 12 summary statistics: the return count and the 11 log-differences of deciles (quantiles at 0.0, 0.1, ..., 1.0) of the non-return distances (including the median). Across four lags, the resulting summary vector has 48 elements. Edge cases (e.g. no non-return events or zero quantile differences) are handled using missing values to avoid computational issues. This construction can be regarded as a hybrid approach, since the observed dataset is not used directly but is instead represented by a set of carefully designed summary statistics that combine return counts with decile-based quantile differences of non-return distances.

We propose a weighted distance metric that combines normalized return-count distances and statistical distances on non-return displacements across lags. Let $\mathcal{D}_1^{(i)}, \ldots, \mathcal{D}_8^{(i)}$ be the distances for the $i$-th simulated dataset, where $\mathcal{D}_{1:4}^{(i)}$ correspond to return counts and $\mathcal{D}_{5:8}^{(i)}$ correspond to statistical distances of non-return displacements at each lag. The combined distance is defined as:
\begin{align*}
\mathcal{D}^{(i)} = \omega \frac{\sum_{k=1}^4 \mathcal{D}^{(i)}_k}{\underset{j=1,\ldots,N}{\text{max}}  \sum_{k=1}^4 \mathcal{D}^{(j)}_k} + (1 - \omega)\frac{\sum_{k=5}^8 \mathcal{D}^{(i)}_k}{\underset{j=1,\ldots,N}{\text{max}} \sum_{k=5}^8 \mathcal{D}^{(j)}_k} \qquad \omega\in[0,1].
\end{align*}
Simulation experiments indicated that lower weights on the return-count component ($\omega \leq 0.2$) yielded more accurate model recovery, suggesting that fine-grained information in the non-return displacements is more informative than return counts alone. All results reported below use $\omega=0.2$.

As a benchmark, we also implemented the ABC-Stat method from \citet{drovandi2022}, using the same 48 summary statistics and a weighted Euclidean distance. The weights are given by the reciprocal of the median absolute deviation (MAD) of the prior predictive distribution for each statistic \citep{prangle2015}. Given observed data show skewness, we implemented ABC-MMD and ABC-Wass with both untrasformed data and log-transformed data.

To assess model recovery, we generated 100 datasets from each model using the parameter values in \cite{marchand2017}: $(\alpha,\gamma,p_0)=(1.7,34,0.6)$ for $M_1$, $(\alpha,\gamma,p_0)=(1.83,46,0.65)$ for $M_2$, and $(\alpha,\gamma,p_0,d_0)=
(1.65,32,0.43,758)$ for $M_3$. Each simulated dataset mirrors the structure of the real data used in Section \ref{sec:real}, with $n_t=66$ individuals observed over $n_d=63$ days. 

Table \ref{ex_toad_table} summarizes the posterior model probabilities across 100 datasets for each method, and Appendix G includes the corresponding confusion matrices and boxplots. ABC-Wass (log) consistently exhibits the highest mean posterior probabilities for the correct models, regardless of which model generated the data. Boxplots in Appendix G further indicate tight interquartile ranges, reflecting consistent performance across replicates. ABC-CvM and ABC-MMD (log) also perform well, although ABC-MMD (log) shows slightly more variability and lower mean probabilities compared to ABC-Wass (log). Overall, Model $M_2$ is the most easily recognizable, while Models $M_1$ and $M_3$ display more variable classification depending on the method.

ABC-Stat demonstrates good performance, with mean posterior probabilities around 0.70–0.75 for Models $M_1$ and $M_3$, and higher for Model $M_2$,  similar to ABC-CvM and ABC-MMD. Parameter estimation is also similar to other methods.

NN exhibits lower classification accuracy across all models, with mean posterior probabilities close to 0.50, indicating near-random performance. Parameter estimation is also poor, with high variability. This may be an indication of a tendency to overfit, so that similar models are less easy to identify. Again, Model $M_2$ shows a higher average posterior probability (around 0.80), although still lower with respect to other methods. In this example, the NN trained using parameters drawn from the prior performs noticeably worse than in the previous hypothesis testing settings (Sections \ref{sub:normal} and \ref{sub:gandk}). While prior-based training improves NN performance in simpler hypothesis testing problems, it does not yield comparable gains here. A likely explanation is the combination of increased model complexity, temporal dependence in the data, and substantial within-model variability induced by prior sampling, together with computational constraints that limit NN training to $10^5$ simulated datasets, making the learning of a stable global decision boundary particularly challenging in this setting.

ABC-SA, as implemented in the \texttt{abctools} R package, encountered memory limitations; to mitigate this, the number of summary statistics was reduced from 48 to 20. This likely contributed to its decreased performance relative to ABC-Stat. Model $M_2$ remains relatively easy to identify, with an average posterior probability larger than 0.95, but Models $M_1$ and $M_3$ are frequently misclassified. Parameter estimation under ABC-SA remains competitive with respect to the other methods.

ABC-QDA generally follows the overall structure of the ABC methods but shows lower classification accuracy, particularly for Model $M_1$, which is often misclassified as Model $M_3$, and viceversa. Parameter estimation is similarly affected, with the highest MSE values, together with NN. This may be an indication that wrong models are estimated as closely as possible to the correct model.

Overall, these results highlight the importance of carefully selecting the distance metric and considering appropriate data transformations. Among all methods, ABC-Wass (log) provides the best performance in model selection, combining high posterior probabilities with low variability, and demonstrates competitive reliability for parameter estimation when models are correctly identified.

\begin{scriptsize}
\begin{table}[h]
\hspace*{-1.5cm} 
\centering
\footnotesize
\begin{tabular}{c|c|cccccc|cc|cc}
      &             & \multicolumn{6}{c|}{\textbf{ABC}}                          & \multicolumn{2}{c|}{\textbf{NN}} & \multicolumn{2}{c}{\textbf{ABC}} \\
      &             & \textbf{CvM} & \textbf{MMD} & \textbf{MMD} & \textbf{Wass} & \textbf{Wass}& \textbf{Stat} & \textbf{fixed} & \textbf{prior} & \textbf{SA}                 &  \textbf{QDA}                 \\
      &             & \multicolumn{2}{c}{}                          & \multicolumn{1}{c}{\textbf{(log)}} & \multicolumn{1}{c}{} & \multicolumn{1}{c}{\textbf{(log)}} & \multicolumn{1}{c|}{}  & \multicolumn{2}{c|}{} & \multicolumn{2}{c}{} \\
      \hline
$M_1$ & $\Pr(M=M_1)$ & 0.711 & 0.714 & 0.802 & 0.784 & 0.926 & 0.741 & 0.471 & 0.504 & 0.632 & 0.478 \\
 & $\hat{\alpha}$ & 1.732 & 1.796 & 1.778 & 1.703 & 1.687 & 1.772 & 1.598 & 0.532 & 1.748 & 1.811 \\
 & $MSE_{\alpha}$ & 0.012 & 0.027 & 0.022 & 0.003 & 0.001 & 0.014 & 0.052 & 1.367 & 0.021 & 0.030 \\
 & $\hat{\gamma}$ & 34.891 & 38.232 & 36.682 & 34.193 & 33.877 & 35.511 & 29.298 & 0.463 & 36.096 & 38.701 \\
 & $MSE_{\gamma}$ & 8.198 & 35.123 & 20.877 & 2.602 & 2.089 & 12.023 & 46.179 & 1124.68 & 19.598 & 41.278 \\
 & $\hat{p}_0$ & 0.609 & 0.657 & 0.648 & 0.602 & 0.587 & 0.638 & 0.523 & 0.508 & 0.619 & 0.667 \\
 & $MSE_{p_0}$ & 0.004 & 0.011 & 0.008 & 0.002 & 0.001 & 0.006 & 0.020 & 0.014 & 0.009 & 0.013 \\
\hline
$M_2$ & $\Pr(M=M_2)$ & 0.958 & 0.918 & 0.968 & 0.951 & 0.989 & 0.957 & 0.781 & 0.621 & 0.962 & 0.881 \\
 & $\hat{\alpha}$ & 1.841 & 1.898 & 1.882 & 1.829 & 1.822 & 1.869 & 1.701 & 0.487 & 1.861 & 1.908 \\
 & $MSE_{\alpha}$ & 0.009 & 0.023 & 0.019 & 0.003 & 0.001 & 0.013 & 0.048 & 1.811 & 0.018 & 0.032 \\
 & $\hat{\gamma}$ & 45.179 & 48.088 & 47.289 & 45.912 & 45.783 & 46.223 & 39.121 & 0.467 & 46.978 & 49.003 \\
 & $MSE_{\gamma}$ & 6.498 & 31.975 & 20.092 & 2.797 & 2.011 & 11.401 & 50.876 & 2073.221 & 18.889 & 43.472 \\
 & $\hat{p}_0$ & 0.638 & 0.691 & 0.677 & 0.653 & 0.641 & 0.672 & 0.581 & 0.498 & 0.661 & 0.698 \\
 & $MSE_{p_0}$ & 0.005 & 0.012 & 0.009 & 0.002 & 0.001 & 0.007 & 0.018 & 0.029 & 0.007 & 0.015 \\
\hline
$M_3$ & $\Pr(M=M_3)$ & 0.701 & 0.719 & 0.881 & 0.732 & 0.909 & 0.738 & 0.458 & 0.482 & 0.652 & 0.501 \\
 & $\hat{\alpha}$ & 1.659 & 1.698 & 1.682 & 1.641 & 1.652 & 1.689 & 1.599 & 0.533 & 1.669 & 1.709 \\
 & $MSE_{\alpha}$ & 0.006 & 0.020 & 0.015 & 0.002 & 0.001 & 0.010 & 0.041 & 1.251 & 0.016 & 0.026 \\
 & $\hat{\gamma}$ & 32.603 & 34.514 & 33.278 & 32.296 & 32.111 & 33.821 & 28.003 & 0.464 & 33.998 & 35.124 \\
 & $MSE_{\gamma}$ & 4.102 & 30.178 & 17.867 & 2.201 & 1.723 & 9.312 & 48.523 & 994.535 & 16.012 & 39.563 \\
 & $\hat{p}_0$ & 0.438 & 0.482 & 0.459 & 0.428 & 0.432 & 0.458 & 0.392 & 0.518 & 0.451 & 0.498 \\
 & $MSE_{p_0}$ & 0.003 & 0.010 & 0.007 & 0.001 & 0.001 & 0.006 & 0.014 & 0.012 & 0.007 & 0.011 \\
 & $\hat{d}_0$ & 762 & 785 & 770 & 757 & 758 & 773 & 710 & 240 & 769 & 800 \\
 & $MSE_{d_0}$ & 8.598 & 32.421 & 19.289 & 2.503 & 1.893 & 10.504 & 49.689 & 57437.857 & 21.179 & 44.120 \\
\end{tabular}
\caption{Results for the toad movement example across 100 repetitions. For each method, the probability of selecting the correct model $\Pr(M = M_i)$, the posterior mean of the parameters for the selected model, and the corresponding MSEs are shown. For all ABC methods, $q\%=0.1\%$ is used.}
\label{ex_toad_table}
\end{table}
\end{scriptsize}

\section{Real Data Example}
\label{sec:real}

The data consist of spatial location recordings of Fowler’s toads collected along the north shore of Lake Erie (Ontario, Canada), a linear sand dune system. Individuals were equipped with radiotransmitters and tracked between mid-June and late August in 2009 and 2010. Locations were recorded during both nocturnal foraging periods and daytime refuge use. In the analysis that follows, we focus exclusively on daytime refuge locations, as these are most informative for characterising return behavior. The dataset comprises observations on $n_t = 66$ individuals over $n_d = 63$ days.

To facilitate modelling, the original two-dimensional coordinates were projected onto a one-dimensional axis aligned with the shoreline. This transformation removes variability orthogonal to the shoreline that is driven by ecological constraints unrelated to return dynamics, while preserving movement along the primary spatial axis of interest. The resulting data can be represented as an $n_d \times n_t$ matrix of projected locations. The full dataset is publicly available at
\url{https://github.com/pmarchand1/fowlers-toad-move/}

We employ the same set of summary statistics introduced in Section~\ref{sub:toad}, which jointly capture return frequency and the distributional properties of non-return movements, yielding a 48-dimensional summary representation. Exploratory analysis of the non-return component reveals pronounced right-skewness and heavy-tailed behaviour across all considered lags. These features motivate the use of monotone transformations when applying full-data ABC discrepancies, particularly for ABC-MMD and ABC-Wass, in order to reduce sensitivity to extreme displacements.

Notably, the posterior support assigned to $M_3$ by the full-data ABC methods exceeds the highest probabilities observed for simulated datasets generated under $M_3$ (Table~\ref{ex_toad_table}, Section~\ref{sub:toad}). One possible explanation is that mild model misspecification amplifies support for the most flexible return mechanism while diminishing evidence for simpler alternatives. This interpretation is consistent with existing results on the robustness of distance-based ABC under misspecification \citep{frazier2020, legramanti2023}; see also Theorem \ref{thm:misspec}.

\begin{table}[h!]
\begin{tabular}{l|ccc}
                                         & \multicolumn{1}{l}{$\Pr(M=M_1)$}                    & \multicolumn{1}{l}{$\Pr(M=M_2)$}                    & \multicolumn{1}{l}{$\Pr(M=M_3)$} \\ \hline
\textbf{ABC-CvM}                         & 0.08                                                & 0.00                                                & 0.92                             \\
\textbf{ABC-MMD}                         & 0.29                                                & 0.00                                                & 0.71                             \\
\textbf{ABC-MMD (log)}                   & 0.07                                                & 0.00                                                & 0.93                             \\
\textbf{ABC-Wass}                        & 0.14                                                & 0.00                                                & 0.86                             \\
\textbf{ABC-Wass (log)}                  & 0.00                                                & 0.00                                                & 1.00                             \\
\textbf{ABC-Stat}                        & 0.40                                                & 0.00                                                & 0.60                             \\
\textbf{ABC-Stat \citep{marchand2017}} & 0.15                                                & 0.00                                                & 0.85                             \\
\textbf{NN - fixed}                              & 0.32 & 0.02                                                & 0.66                             \\
\textbf{NN - with prior}                              & 0.85 &  0.01  & 0.14                            \\
\textbf{ABC-SA}                          & 0.36 & 0.00                                                & 0.64                             \\
\textbf{ABC-QDA}                         & 0.13 &  0.09 & 0.79                            
\end{tabular}
\caption{Estimated posterior probabilities of the three toad movement models, comparing different methods with the original results of \cite{marchand2017}.}
\label{tab:toad_table}
\end{table}
 
While our findings broadly reproduce the main empirical patterns reported by \citet{marchand2017}, they differ from their final conclusion that the random return model $M_1$ provides the best description of the data. That conclusion relies in part on the limited performance of their ABC procedure in simulation, which prompts the use of auxiliary criteria based on refuge counts. In contrast, the full-data ABC methods considered here demonstrate strong and stable performance in simulation (Section~\ref{sub:toad}), lending greater confidence to the identification of $M_3$. 

Taken together, these results suggest that a distance-dependent return mechanism offers a more coherent explanation of the observed movement patterns than the competing models. Under $M_3$, long-distance movements are less likely to be followed by returns to previously visited sites, whereas under $M_1$ and $M_2$ the return probability remains constant irrespective of distance. This behavior is ecologically plausible, as establishing new refuges may be more likely when individuals move far from prior locations. These conclusions should nevertheless be interpreted cautiously, as no formal goodness-of-fit analysis was conducted.

\begin{figure}[h!]
\hspace*{-0.6cm} 
\includegraphics[scale=0.28]{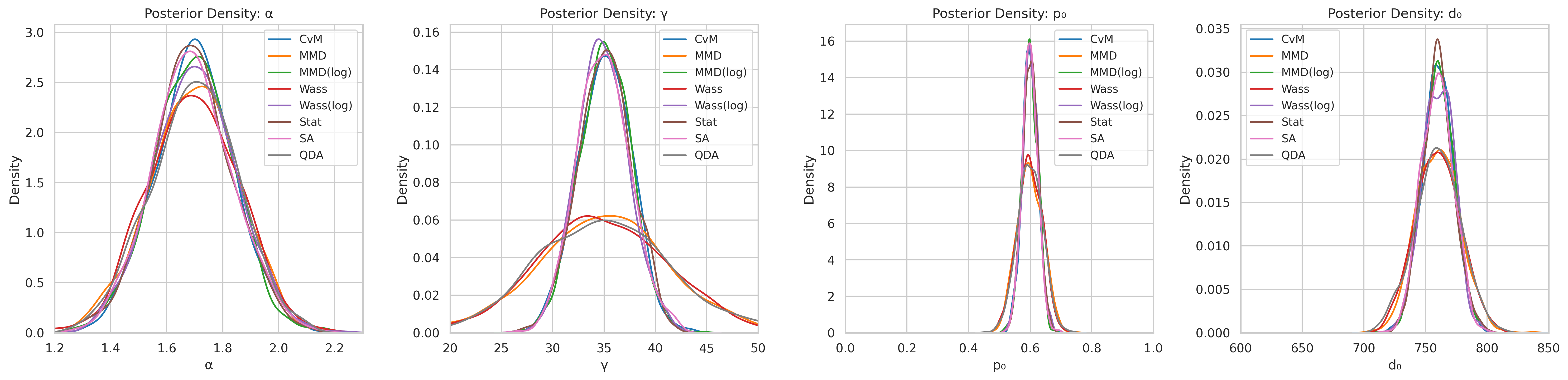}
\caption{Posterior distributions of model $M_3$ parameters across the considered methods.}
\label{ex_toad_posteriors}
\end{figure}

Turning to parameter inference (Figure \ref{ex_toad_posteriors}), all ABC-based methods yield stable estimates of the return strength parameter $\alpha$. Greater uncertainty is observed for $\gamma$, $p_0$, and $d_0$, particularly for ABC-MMD, ABC-Wass, and ABC-QDA.
For clarity, NN results are omitted from Figure~\ref{ex_toad_posteriors}: when trained on fixed parameter values, the NN effectively learns a point-mass mapping rather than a posterior distribution, while prior-based training produces highly variable and strongly biased estimates relative to the ABC methods.
Overall, once the correct model is identified, distance-based ABC approaches provide reliable parameter inference in this application.

\section{Conclusion}
\label{sec:conclu}

This work investigated approximate Bayesian computation (ABC) as a tool for model selection in settings with intractable likelihoods. While ABC is widely used for parameter inference, its application to model choice remains challenging, primarily due to sensitivity to the choice of summary statistics. When summaries are insufficient, substantial information loss can lead to unreliable posterior model probabilities.

The theoretical results presented in Section~\ref{sec:ABCmodelchoice} establish both consistency of Wasserstein-based ABC under correct specification and robustness under model misspecification. The assumptions required are mild and easily verifiable, particularly in the common i.i.d. setting, and empirical results suggest that good performance can extend to more complex dependence structures. 

The theoretical results are supported by an extensive empirical evaluation of ABC model selection based on statistical distances computed on the full data. Through a range of simulation studies, we showed that full data ABC approaches can closely match ABC with sufficient statistics when such statistics are available, and can otherwise provide strong support for the correct model in more challenging settings. These results suggest that discrepancy-based ABC methods preserve key distributional features necessary for accurate likelihood-free model comparison.

Across all simulations, the Wasserstein distance emerged as the most reliable discrepancy for model selection, achieving faster convergence to the true model, lower misclassification rates, and stable posterior model probabilities. The Cramér–von Mises distance and the Maximum Mean Discrepancy also performed well in simpler settings, although their effectiveness depended on sample size, model complexity, and appropriate data transformations. These findings align with the previous results of \cite{drovandi2022} about parameter inference and highlight the importance of tailoring the discrepancy measure to the data-generating mechanism.

Across the simulation studies and the real-data application, the comparison with neural network–based approaches highlights a fundamental difference between discriminative and simulation-based likelihood-free inference. When trained using parameters drawn from the prior, NN performance can improve in simple hypothesis testing problems, particularly under a point null hypothesis, where prior integration sharpens the marginal evidence under the alternative. However, in more complex settings, including general model selection problems, NN-based inference proves sensitive to prior-induced variability, limited training data, and model complexity, leading to unstable posterior model probabilities and biased parameter estimates. In contrast, ABC methods based on statistical distances consistently provide well-calibrated posterior model probabilities and coherent parameter inference, underscoring their robustness and suitability for likelihood-free Bayesian model choice.

While this study employed the standard rejection ABC algorithm to enable fair comparisons across distance metrics, integrating statistical distances into more efficient ABC schemes \citep{toni2010, didelot2011} represents a natural next step, particularly in settings where computational cost is critical. Rejection ABC also allows full parallelisation, which we leveraged in our computational performance evaluations. Although more advanced ABC algorithms may improve efficiency, we do not expect them to substantially alter the comparative performance of the distance metrics.

Although full data ABC methods avoid low-dimensional summary statistics, they remain subject to the curse of dimensionality: Wasserstein distances grow exponentially with dimension \citep{fournier2015rate, bernton2019}, and MMD may lose power due to kernel sensitivity \citep{ramdas2015decreasing}. Effective application therefore requires careful choice of distance metric and strategies to manage computational complexity, such as dimensionality reduction, adaptive weighting, or feature learning. In other words, full data ABC does not eliminate the curse of dimensionality, it shifts the challenge, requiring practitioners to regularize the discrepancy measure and manage high-dimensional effects. 

In summary, our findings demonstrate that ABC with statistical distances offers a promising framework for likelihood-free model selection. These methods provide accurate, interpretable results, reduce reliance on summary statistics, and perform robustly under realistic model misspecification. 




\section*{Appendix A: ABC algorithms}

This Appendix presents two rejection ABC algorithms used for parameter inference and one used for model selection. 

Algorithms \ref{alg:abc_sampler} and \ref{alg:full_data_abc} are aimed at statistical inference. Algorithm \ref{alg:abc_sampler} makes use of summary statistics, while Algorithm \ref{alg:full_data_abc} makes use of statistical distances. Algorithm \ref{alg:abc-mc} performs model selection using summary statistics and it has been shown to be inconsistent in most cases by \cite{robert2011}. 

\begin{algorithm}
 \caption{Summary-based ABC}\label{alg:abc_sampler}
 \begin{algorithmic}[1]
 \State \textbf{Input:} Given a set of observations $\bm{y}=(y_1, \ldots, y_n)^T$, an observed summary statistics $\bm{\eta}(\bm{y})$, a tolerance level $\varepsilon$, and a distance $\rho(\cdot, \cdot)$:
 \For{$i = 1, \dots, N$}
 \Repeat
     \State Generate $\bm{\theta}^*$ from $\pi(\bm{\theta})$.
     \State Generate $\bm{z} = (z_1, z_2, ..., z_n)^T$ from $ P_{\bm{\theta}^*}^{(n)}$.
 \Until{$\rho(\bm{\eta}(\bm{y}), \bm{\eta}(\bm{z})) \leq \varepsilon$}
 \State Set $\bm{\theta}^{(i)} = \bm{\theta}^*$.
 \EndFor
 \State \textbf{Output:} A set of values $(\bm{\theta}^{(1)}, \ldots, \bm{\theta}^{(N)})$ from $\pi_\varepsilon (\bm{\theta} | \bm{\eta}(\bm{y}))$.
 \end{algorithmic}
\end{algorithm}

\begin{algorithm}
\caption{Discrepancy-based ABC}\label{alg:full_data_abc}
 \begin{algorithmic}[1]  
  \State \textbf{Input:} Given a set of observations $\bm{y}=(y_1, \ldots, y_n)^T$, a tolerance level $\varepsilon$, and a discrepancy metric $\mathcal{D}(\cdot, \cdot)$:
  \For{$i = 1, \dots, N$}
  \Repeat
    \State Generate $\bm{\theta}^*$ from $\pi(\bm{\theta})$. 
    \State Generate $\bm{z} = (z_1, z_2, ..., z_n)^T$ from $ P_{\bm{\theta}^*}^{(n)}$.
    \Until{$\mathcal{D}(\hat{\mu}_{\bm{\theta}_0}, \hat{\mu}_{\bm{\theta}^*}) \leq \varepsilon$}
    \State Set $\bm{\theta}^{(i)} = \bm{\theta}^*$.
  \EndFor
 \State \textbf{Output:} A set of values $(\bm{\theta}^{(1)}, \ldots, \bm{\theta}^{(N)})$ from $\pi_\varepsilon (\bm{\theta} | \bm{y})$.
 \end{algorithmic}
\end{algorithm}

\begin{algorithm}
 \caption{Summary-based ABC-MC}\label{alg:abc-mc}
 \begin{algorithmic}[1]
 \State \textbf{Input:} Given a set of observations $\bm{y}=(y_1, \ldots, y_n)^T$, a set of possible models $\{M_1, M_2, \ldots, M_K\}$, an observed summary statistics $\bm{\eta}(\bm{y})$, a tolerance level $\varepsilon$, and a distance $\rho(\cdot, \cdot)$:
 \For{$i = 1, \dots, N$}
  \Repeat
     \State Generate $M_{k^*}$ from $\pi(M_k)$, $k=1, \ldots,K$.
     \State Generate $\bm{\theta}_{k^*}$ from $\pi_{k^*}(\bm{\theta}_{k^*})$.
     \State Generate $\bm{z} = (z_1, ..., z_n)^T$ from $P_{M_{k^*}, \bm{\theta}_{k^*}}^{(n)}$.
 \Until{$\rho(\bm{\eta}(\bm{y}), \bm{\eta}(\bm{z})) \leq \varepsilon$},
 \State Set $M^{(i)} = M_{k^*}$ and $\bm{\theta}^{(i)} = \bm{\theta}_{k^*}$.
 \EndFor
 \State \textbf{Output:} A set of values $(M^{(1)}, \ldots, M^{(N)})$ and $(\bm{\theta}^{(1)}, \ldots, \bm{\theta}^{(N)})$ from $\pi_{\varepsilon}(M_k|\bm{\eta}(\bm{y}))$ and $\pi_{\varepsilon,k} (\bm{\theta}_k | \bm{\eta}(\bm{y}),M_k)$, for $k=1, \ldots, K$, respectively.
 \end{algorithmic}
\end{algorithm}

\newpage

\section*{Appendix B: Proof of Theorem 3.1}

\begin{proof}
Let the observed data $\mathbf{y}$ be i.i.d. from $P^{(n)}_{M_0,\bm{\theta}_0}$. For a candidate model $M_{k^*}$ with parameter $\bm{\theta}_{k^*}\in\Theta_{k^*}$ denote the population law by $P^{(n)}_{M_{k^*},\bm{\theta}_{k^*}}$. Define the ABC numerator for model $M_{k^*}$ with threshold $\varepsilon_n$ by
\begin{equation}
N_{k^*}(n):=\pi(M_{k^*})\int_{\Theta_{k^*}}\pi_{M_{k^*}}(\bm{\theta}_{k^*})\,
\mathbb P\big(\mathcal{W}_p(\hat\mu_{0,\bm{\theta}_0},\hat\mu_{{k^*},\bm{\theta}_{k^*}})\le\varepsilon_n\big)\,d\bm{\theta}_{k^*},
\label{eq:nkn}
\end{equation}
and the ABC-Wass model-posterior by $\pi_{\varepsilon_n}(M_{k^*}\mid\mathbf y)=N_{k^*}(n)/\sum_{j}N_j(n)$.

We prove that, under Assumptions (A1)--(A4) and for any sequence $\varepsilon_n\to0$ with eventually $\varepsilon_n<\min_{k\ne0}\delta_{k^*}/2$, one has
\[
\pi_{\varepsilon_n}(M_0\mid\mathbf y)\xrightarrow{p}1.
\]

The proof has two parts: (i) $N_0(n)$ is bounded away from zero in probability; (ii) for every $k^*\neq0$, $N_{k^*}(n)\xrightarrow{p}0$. Together these imply posterior concentration on $M_0$.

\paragraph{(i) Lower bound for $N_0(n)$.}
Define
\[
q_{0,n}:=\mathbb P\big(\mathcal{W}_p(\hat{\mu}_{0},\hat\mu_{k^*,\bm{\theta}_{k^*}})\le\varepsilon_n\big).
\]
By the triangle inequality,
\[
\mathcal{W}_p(\hat{\mu}_{0},\hat\mu_{k^*,\bm{\theta}_{k^*}})
\le \mathcal{W}_p(\hat{\mu}_{0},P_{M_0,\bm{\theta}_0})+\mathcal{W}_p(\hat\mu_{k^*,\bm{\theta}^*},P_{M_0,\bm{\theta}_0}).
\]
Hence, for $a\in(0,\varepsilon_n)$,
\[
\{\mathcal{W}_p(\hat{\mu}_{0},\hat\mu_{k^*,\bm{\theta}_{k^*}})>\varepsilon_n\}
\subseteq
\{\mathcal{W}_p(\hat{\mu}_{0},P_{M_0,\bm{\theta}_0})>a\}\cup\{\mathcal{W}_p(\hat\mu_{k^*,\bm{\theta}_{k^*}},P_{M_0,\bm{\theta}_0})>\varepsilon_n-a\}.
\]
It is possible to choose $a=\varepsilon_n/2$. Then
\[
1-q_{0,n}\le\mathbb P\big(\mathcal{W}_p(\hat{\mu}_{0},P_{M_0,\bm{\theta}_0})>\tfrac{\varepsilon_n}{2}\big)
+ \mathbb P\big(\mathcal{W}_p(\hat\mu_{k^*,\bm{\theta}_{k^*}},P_{M_0,\bm{\theta}_0})>\tfrac{\varepsilon_n}{2}\big).
\]
By Assumption (A4), $\mathcal{W}_p(\hat{\mu}_{0},P_{M_0,\bm{\theta}_0})\xrightarrow{p}0$. For the second term $\mathbb P\big(\mathcal{W}_p(\hat\mu_{k^*,\bm{\theta}_{k^*}},P_{M_0,\bm{\theta}_0})>\tfrac{\varepsilon_n}{2}\big)$, we can use the Markov's inequality
\[
\mathbb P\big(\mathcal{W}_p(\hat\mu_{k^*,\bm{\theta}_{k^*}},P_{M_0,\bm{\theta}_0})>\tfrac{\varepsilon_n}{2}\big)
\le \frac{2\mathbb E[\mathcal{W}_p(\hat\mu_{k^*,\bm{\theta}_{k^*}},P_{M_0,\bm{\theta}_0})]}{\varepsilon_n}
\le \frac{2\Delta_n}{\varepsilon_n}\xrightarrow{n\to\infty}0,
\]
because, under Assumption (A4), $\Delta_n := \sup_{M_{k^*}, \bm{\theta}_{k^*}} \mathbb{E}[\mathcal{W}_p(\hat{\mu}_{k^*,\bm{\theta}_{k^*}}, P^{(n)}_{M_k^*,\bm{\theta}_{k^*}})]= o(\varepsilon_n)$. Thus $q_{0,n}\xrightarrow{p}1$.

Let $U$ be a neighbourhood of $\bm{\theta}_0$ with prior mass $\pi_{M_0}(U)>0$. By continuity of $\bm{\theta}_{k^*}\mapsto P_{M_0,\bm{\theta}_{k^*}}$ in $\mathcal{W}_p$, the same argument shows $\mathbb{P}\big(\mathcal{W}_p(\hat{\mu}_{0},\hat\mu_{k^*,\bm{\theta}_{k^*}})\le\varepsilon_n\big) \to 1$ uniformly for $\bm{\theta}_{k^*}\in U$. Hence, by dominated convergence,
\[
\int_U \pi_{M_0}(\bm{\theta}_{k^*})\,\mathbb{P}\big(\mathcal{W}_p(\hat{\mu}_{0},\hat\mu_{k^*,\bm{\theta}_{k^*}})\le\varepsilon_n\big)\,d\bm{\theta}_{k^*}
\xrightarrow{p}\pi_{M_0}(U)>0,
\]
and therefore $N_0(n)$ is bounded away from zero in probability, since $\pi(M_0)>0$ by Assumption (A2).

\paragraph{(ii) Upper bound for $N_{k^*}(n)$ when $k^*\neq0$.}
Fix $k^*\neq0$ and let $\delta_{k^*}>0$ be as in Assumption (A1), so that
$\mathcal{W}_p(P^{(n)}_{M_0,\bm{\theta}_0},P^{(n)}_{M^*,\bm{\theta}_{k^*}})\ge \delta_{k^*}$ for all $\bm{\theta}_{k^*}\in\Theta_{k^*}$.
For any $\bm{\theta}_{k^*}$, by the reverse form of the triangle inequality, we get:
\[
\begin{aligned}
\mathcal{W}_p(\hat\mu_{0,\bm{\theta}_0},\hat\mu_{k^*,\bm{\theta}_{k^*}})
&\ge \mathcal{W}_p(\hat\mu_{0,\bm{\theta}_0}, P^{(n)}_{M_{k^*},\bm{\theta}_{k^*}}) - \mathcal{W}_p(\hat\mu_{k^*,\bm{\theta}_{k^*}},P^{(n)}_{M_{k^*},\bm{\theta}_{k^*}}) \\
&\ge \mathcal{W}_p(P^{(n)}_{M_{k^*},\bm{\theta}_{k^*}},P^{(n)}_{M_0,\bm{\theta}_0}) - \mathcal{W}_p(\hat\mu_{0,\bm{\theta}_0}, P^{(n)}_{M_{0},\bm{\theta}_{0}}) - \mathcal{W}_p(\hat\mu_{k^*,\bm{\theta}_{k^*}},P^{(n)}_{M_{k^*},\bm{\theta}_{k^*}}).
\end{aligned}
\]
Hence, for Assumption (A1)
$$
\mathcal{W}_p(\hat\mu_{0,\bm{\theta}_0},\hat\mu_{k^*,\bm{\theta}_{k^*}}) \geq \delta_{k^*} - \mathcal{W}_p(\hat\mu_{0,\bm{\theta}_0}, P^{(n)}_{M_{0},\bm{\theta}_{0}}) - \mathcal{W}_p(\hat\mu_{k^*,\bm{\theta}_{k^*}},P^{(n)}_{M_{k^*},\bm{\theta}_{k^*}}).
$$

We are interested in the event 
$$
\mathcal{W}_p(\hat\mu_{0,\bm{\theta}_0},\hat\mu_{k^*,\bm{\theta}_{k^*}}) \leq \varepsilon_n,
$$
which is equivalent to considering
$$
\mathcal{W}_p(\hat\mu_{0,\bm{\theta}_0}, P^{(n)}_{M_{0},\bm{\theta}_{0}}) + \mathcal{W}_p(\hat\mu_{k^*,\bm{\theta}_{k^*}},P^{(n)}_{M_{k^*},\bm{\theta}_{k^*}}) \geq \delta_{k^*} - \varepsilon_n.
$$

Without loss of generality, select $\varepsilon_n < \delta_{k^*}/2$, then $\delta_{k^*} - \varepsilon_n > \delta_k/2$, and
\[
\Big\{\,\mathcal{W}_p(\hat\mu_{0,\bm{\theta}_0},\hat\mu_{k^*,\bm{\theta}_{k^*}}) \leq \varepsilon_n\,\Big\}
\;\subseteq\;
\Big\{\,\mathcal{W}_p(\hat\mu_{0,\bm{\theta}_0}, P^{(n)}_{M_{0},\bm{\theta}_{0}}) + \mathcal{W}_p(\hat\mu_{k^*,\bm{\theta}_{k^*}},P^{(n)}_{M_{k^*},\bm{\theta}_{k^*}}) \ge \delta_{k^*}/2\,\Big\}.
\]
By the union bound, we obtain
\[
\mathbb P\!\Big(\mathcal{W}_p(\hat\mu_{0,\bm{\theta}_0},\hat\mu_{k^*,\bm{\theta}_{k^*}}) \leq \varepsilon_n\Big)
\;\le\;
\mathbb P\!\Big(\mathcal{W}_p(\hat\mu_{0,\bm{\theta}_0}, P^{(n)}_{M_{0},\bm{\theta}_{0}})\ge \tfrac{\delta_{k^*}}{4}\Big)
\;+\;
\mathbb P\!\Big(\mathcal{W}_p(\hat\mu_{k^*,\bm{\theta}_{k^*}},P^{(n)}_{M_{k^*},\bm{\theta}_{k^*}})\ge \tfrac{\delta_{k^*}}{4}\Big).
\]
The first term on the right hand side tends to $0$ by Assumption (A4), since $\mathcal{W}_p(\hat\mu_{0,\bm{\theta}_0}, P^{(n)}_{M_{0},\bm{\theta}_{0}})\xrightarrow{p}0$.
For the second term, the Markov's inequality and the definition of $\Delta_n$ in Assumption (A4) give 
\[
\mathbb P\!\Big(\mathcal{W}_p(\hat\mu_{k^*,\bm{\theta}_{k^*}},P^{(n)}_{M_{k^*},\bm{\theta}_{k^*}})\ge \tfrac{\delta_{k^*}}{4}\Big)
\;\le\;
\frac{4}{\delta_{k^*}}\,\mathbb E\!\Big[\mathcal{W}_p(\hat\mu_{k^*,\bm{\theta}_{k^*}},P^{(n)}_{M_{k^*},\bm{\theta}_{k^*}})\Big]
\;\le\; \frac{4\Delta_n}{\delta_{k^*}}
\;\xrightarrow{n\to\infty}\;0,
\]
uniformly over $\bm{\theta}_{k^*}\in\Theta_{k^*}$. Therefore $N_{k^*}(n)\xrightarrow{p}0$.

From part (i) there exists $c>0$ such that $\mathbb P(N_0(n)>c)\to1$, while from part (ii) $N_{k^*}(n)\xrightarrow{p}0$ for each $k^*\neq0$. Thus $\sum_{k\neq0}N_k(n)\xrightarrow{p}0$, and consequently
\[
\pi_{\varepsilon_n}(M_0\mid\mathbf y)=\frac{N_0(n)}{N_0(n)+\sum_{k\neq0}N_k(n)}\xrightarrow{p}1,
\]
as required.
\end{proof}

\newpage

\section*{Appendix C: Proof of Theorem 3.2}

\begin{proof}
Consider the misspecified setting where the true data distribution $P^{(n)}_{M_0,\bm{\theta}_0}$ does not belong to any candidate model $M_k \in \mathcal{M}$. Let $(M^\dagger, \bm{\theta}^\dagger)$ denote the model–parameter pair that minimizes the Wasserstein distance to the true distribution:
\[
(M^\dagger, \bm{\theta}^\dagger) = \arg\min_{M_k \in \mathcal{M}, \bm{\theta}_k \in \Theta_k} \mathcal{W}_p(P^{(n)}_{M_0,\bm{\theta}_0}, P^{(n)}_{M_k, \bm{\theta}_k}).
\]

By Assumption (A4), for large $n$ we have
\[
\mathcal{W}_p(\hat{\mu}_{0,\bm{\theta}_0}, P^{(n)}_{M_0,\bm{\theta}_0}) \le \eta^{(n)}, 
\quad 
\mathcal{W}_p(\hat{\mu}_{k,\bm{\theta}_k}, P^{(n)}_{M_k,\bm{\theta}_k}) \le \eta^{(n)}_k,
\]
with $\eta^{(n)}, \eta^{(n)}_k \to 0$ as $n \to \infty$, and for $k=1, \ldots, K$.

By the triangle inequality, for any $(M_k, \bm{\theta}_k)$,
\[
\mathcal{W}_p(\hat{\mu}_{0,\bm{\theta}_0}, \hat{\mu}_{k,\bm{\theta}_k}) 
\ge \mathcal{W}_p(P^{(n)}_{M_0,\bm{\theta}_0}, P^{(n)}_{M_k,\bm{\theta}_k}) - \eta^{(n)} - \eta^{(n)}_k.
\]

Consider first the case of models far from the best approximation.
Let $\delta_k := \mathcal{W}_p(P^{(n)}_{M_0,\bm{\theta}_0}, P^{(n)}_{M_k,\bm{\theta}_k}) - \mathcal{W}_p(P^{(n)}_{M_0,\bm{\theta}_0}, P^{(n)}_{M^\dagger,\bm{\theta}^\dagger}) > 0$. Then for sufficiently large $n$, setting the ABC threshold $\varepsilon_n < \delta_k/2$ ensures
\[
\mathcal{W}_p(\hat{\mu}_{0,\bm{\theta}_0}, \hat{\mu}_{k,\bm{\theta}_k}) \ge \delta_k - \eta^{(n)} - \eta^{(n)}_k > \varepsilon_n.
\]
Hence, the ABC acceptance probability for such $(M_k, \bm{\theta}_k)$ vanishes:
\[
\mathbb{P}\Big(\mathcal{W}_p(\hat{\mu}_{0,\bm{\theta}_0}, \hat{\mu}_{k,\bm{\theta}_k}) \le \varepsilon_n\Big) \to 0.
\]

Next, consider models near the best approximation. Consider a Wasserstein neighborhood of the optimal pair:
\[
\mathcal{N}_\delta := \big\{(M_k, \bm{\theta}_k) : \mathcal{W}_p(P^{(n)}_{M_0,\bm{\theta}_0}, P^{(n)}_{M_k,\bm{\theta}_k}) \le \mathcal{W}_p(P^{(n)}_{M_0,\bm{\theta}_0}, P^{(n)}_{M^\dagger,\bm{\theta}^\dagger}) + \delta \big\}.
\]
For sufficiently small $\delta$ and large $n$, by the triangle inequality and empirical convergence,
\[
\mathcal{W}_p(\hat{\mu}_{0,\bm{\theta}_0}, \hat{\mu}_{k,\bm{\theta}_k}) 
\le \mathcal{W}_p(P^{(n)}_{M_0,\bm{\theta}_0}, P^{(n)}_{M_k,\bm{\theta}_k}) + \eta^{(n)} + \eta^{(n)}_k
\le \mathcal{W}_p(P^{(n)}_{M_0,\bm{\theta}_0}, P^{(n)}_{M^\dagger,\bm{\theta}^\dagger}) + \delta + o(1),
\]
so that for small $\varepsilon_n>\delta$,
\[
\mathbb{P}\Big(\mathcal{W}_p(\hat{\mu}_{0,\bm{\theta}_0}, \hat{\mu}_{k,\bm{\theta}_k}) \le \varepsilon_n \Big) \to 1.
\]

Combining these two results, the ABC-Wass posterior concentrates on the neighborhood $\mathcal{N}_\delta$:
\[
\pi_{\varepsilon_n}(\mathcal{N}_\delta \mid \bm{y}) \overset{p}{\to} 1
\quad \text{as } n \to \infty.
\]
This establishes that, under model misspecification, the ABC posterior asymptotically concentrates on the model–parameter pair \((M^\dagger, \bm{\theta}^\dagger)\) that best approximates the true distribution in Wasserstein distance.\end{proof}

\newpage

\section*{Appendix D: Normal Mean Hypothesis Test}

\begin{figure}[h!]
\centering
\includegraphics[scale=0.4]{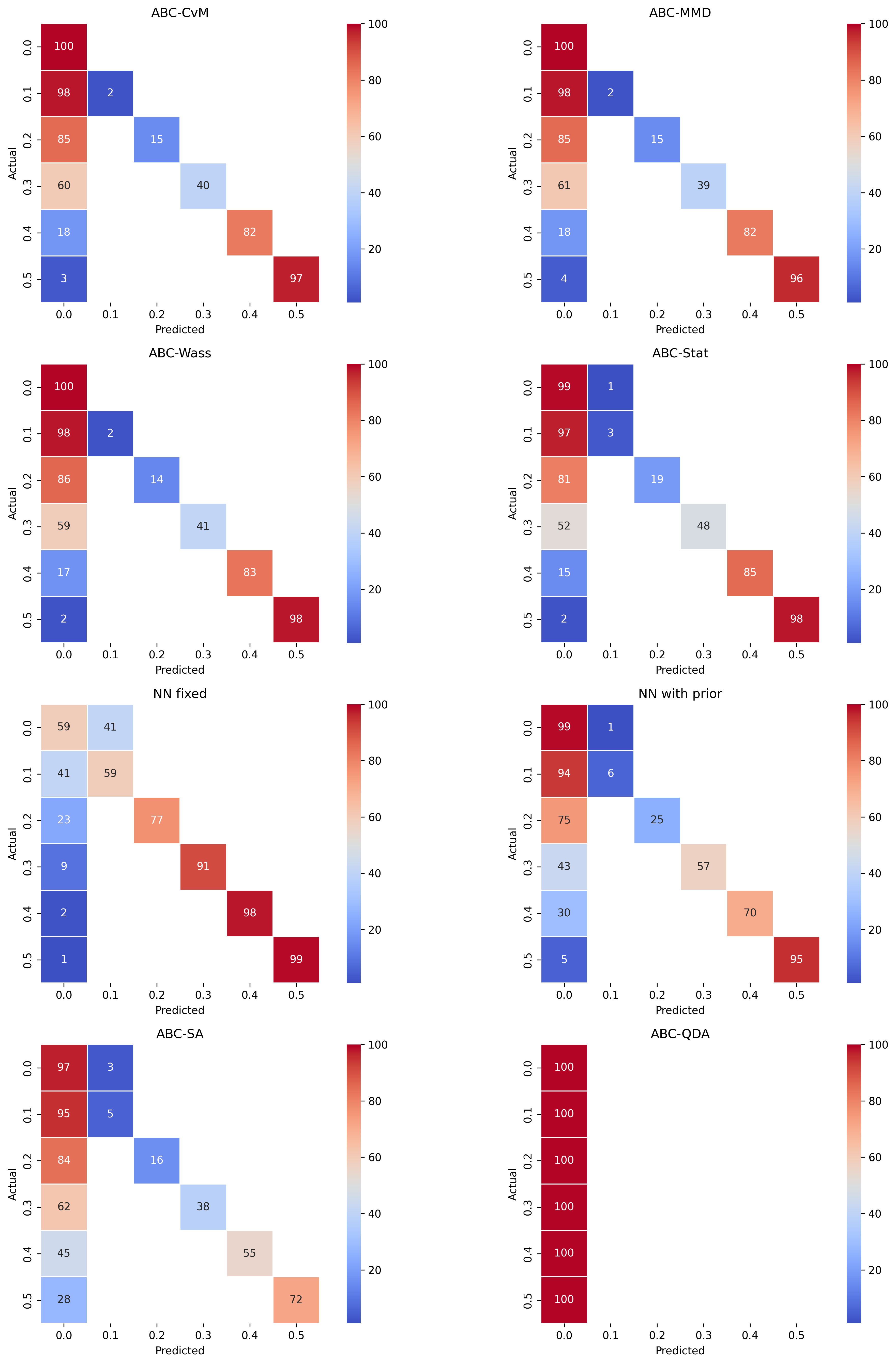}
\caption{Confusion matrices showing model selection performance for each method in the normal example with $n=100$.}
\label{ex_normal_CM}
\end{figure}
\begin{figure}
\includegraphics[scale=0.5]{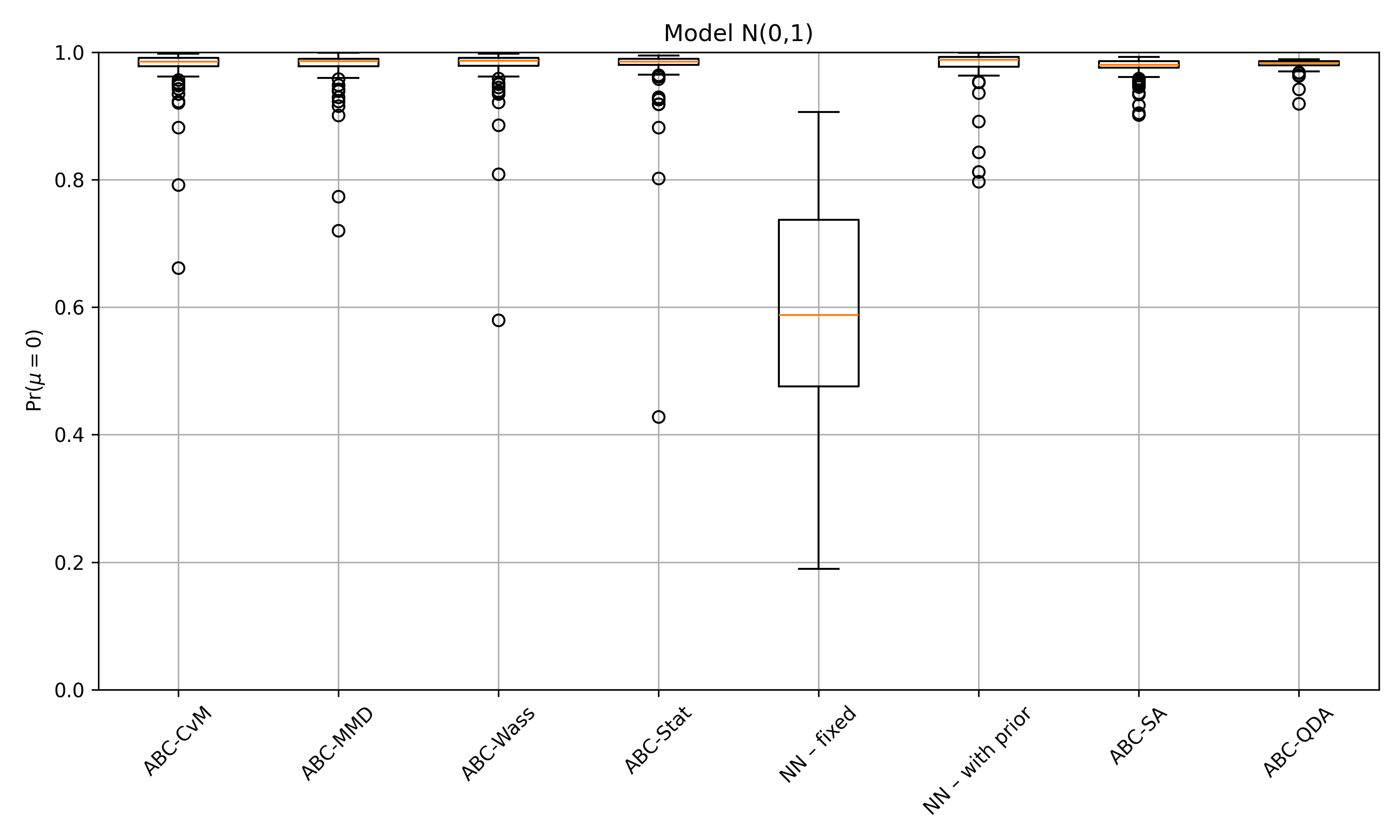}
\caption{Boxplots of the estimated posterior probability of $H_0$ across 100 datasets generated from $N(0,1)$.}
\label{fig:normal_boxplots0}
\end{figure}
\begin{figure}
\includegraphics[scale=0.5]{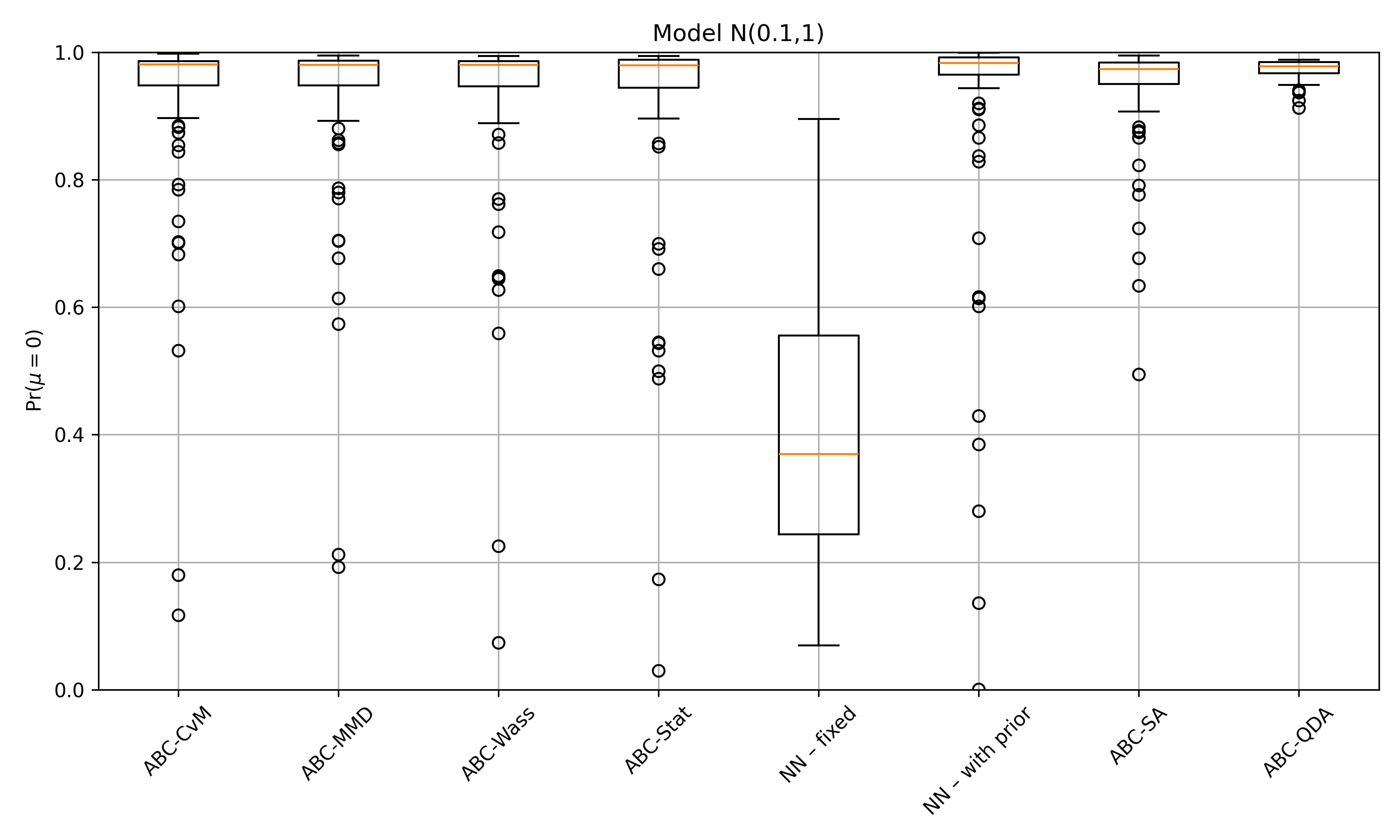}
\caption{Boxplots of the estimated posterior probability of $H_0$ across 100 datasets generated from $N(0.1,1)$.}
\label{fig:normal_boxplots1_1}
\end{figure}
\begin{figure}
\includegraphics[scale=0.5]{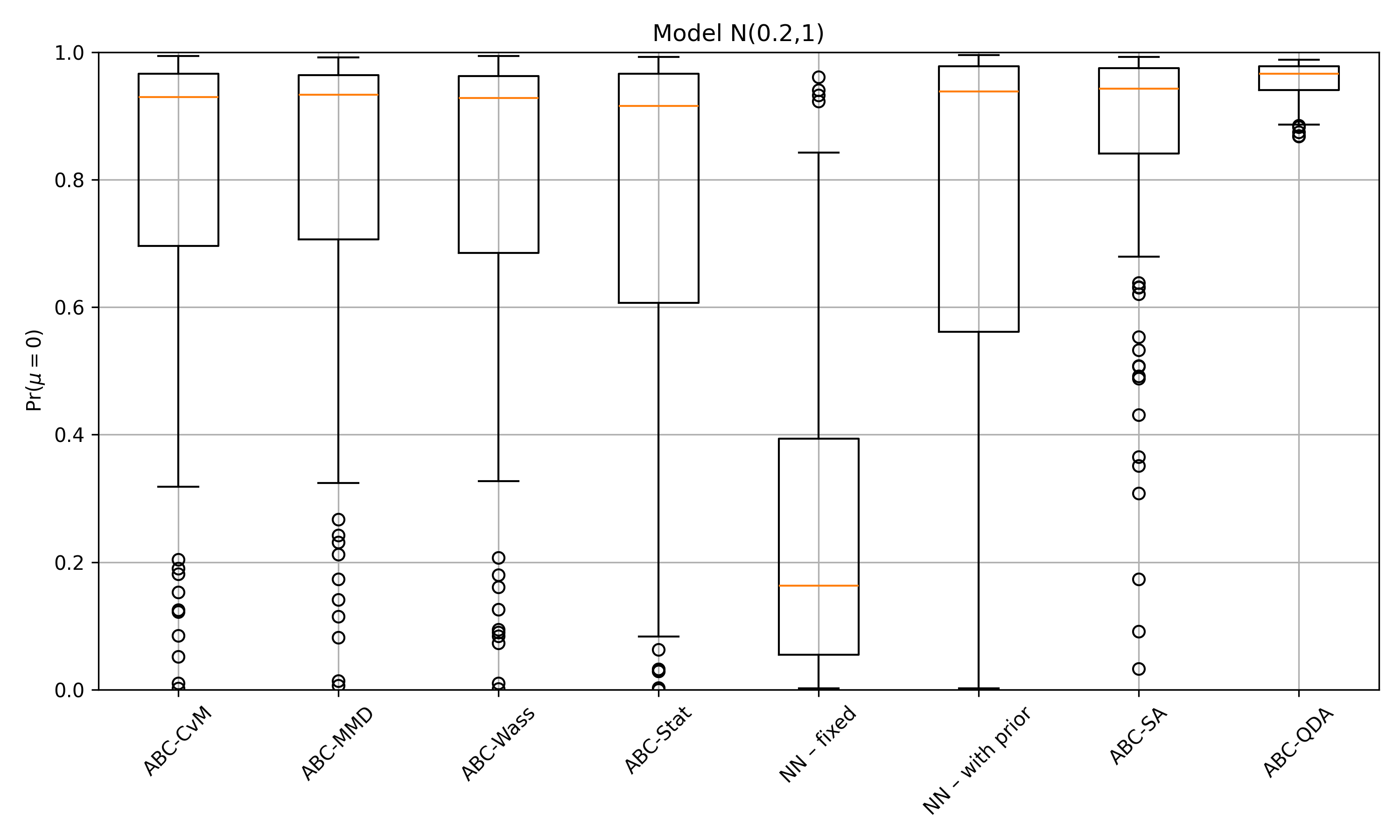}
\caption{Boxplots of the estimated posterior probability of $H_0$ across 100 datasets generated from $N(0.2,1)$.}
\label{fig:normal_boxplots1_2}
\end{figure}
\begin{figure}
\includegraphics[scale=0.5]{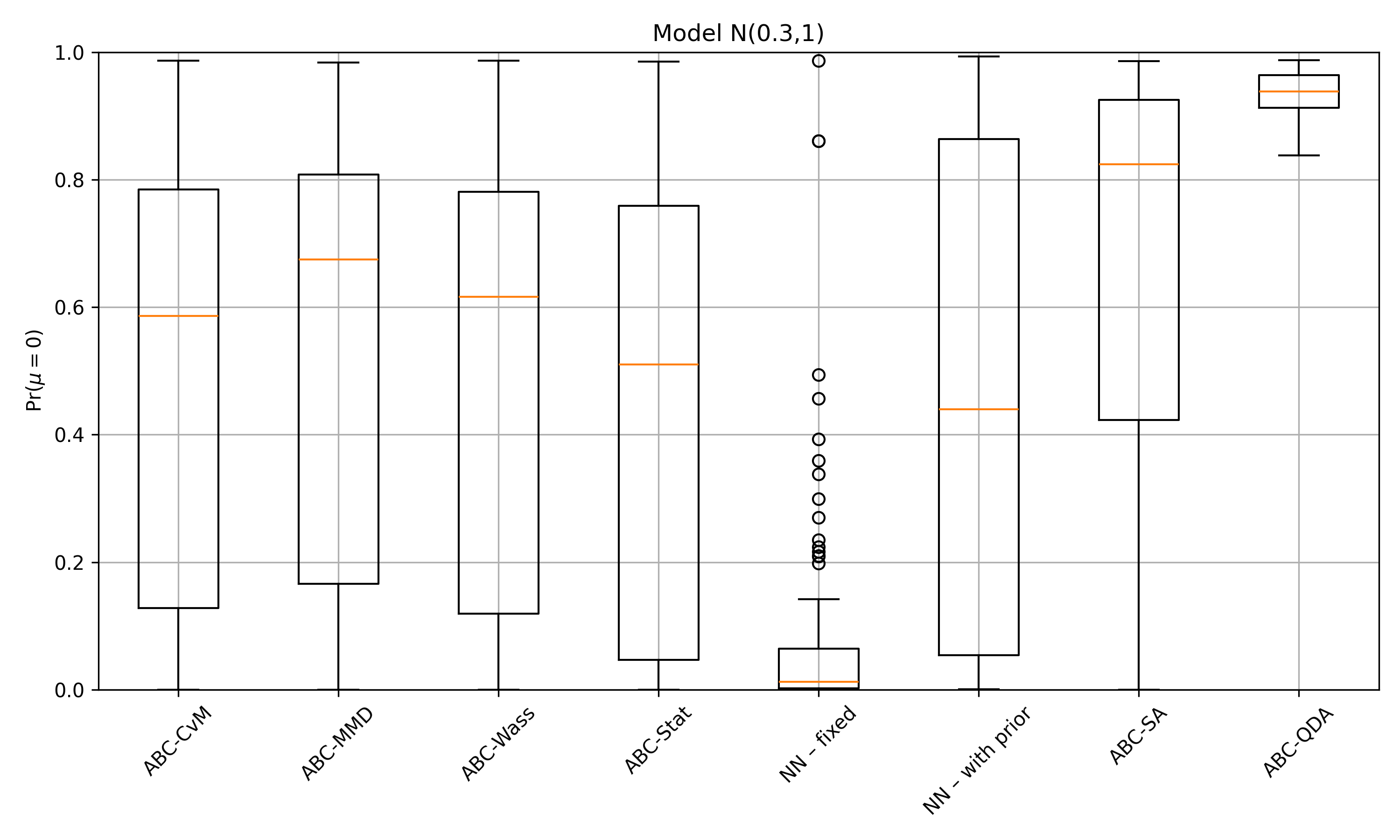}
\caption{Boxplots of the estimated posterior probability of $H_0$ across 100 datasets generated from $N(0.3,1)$.}
\label{fig:normal_boxplots1_3}
\end{figure}
\begin{figure}
\includegraphics[scale=0.5]{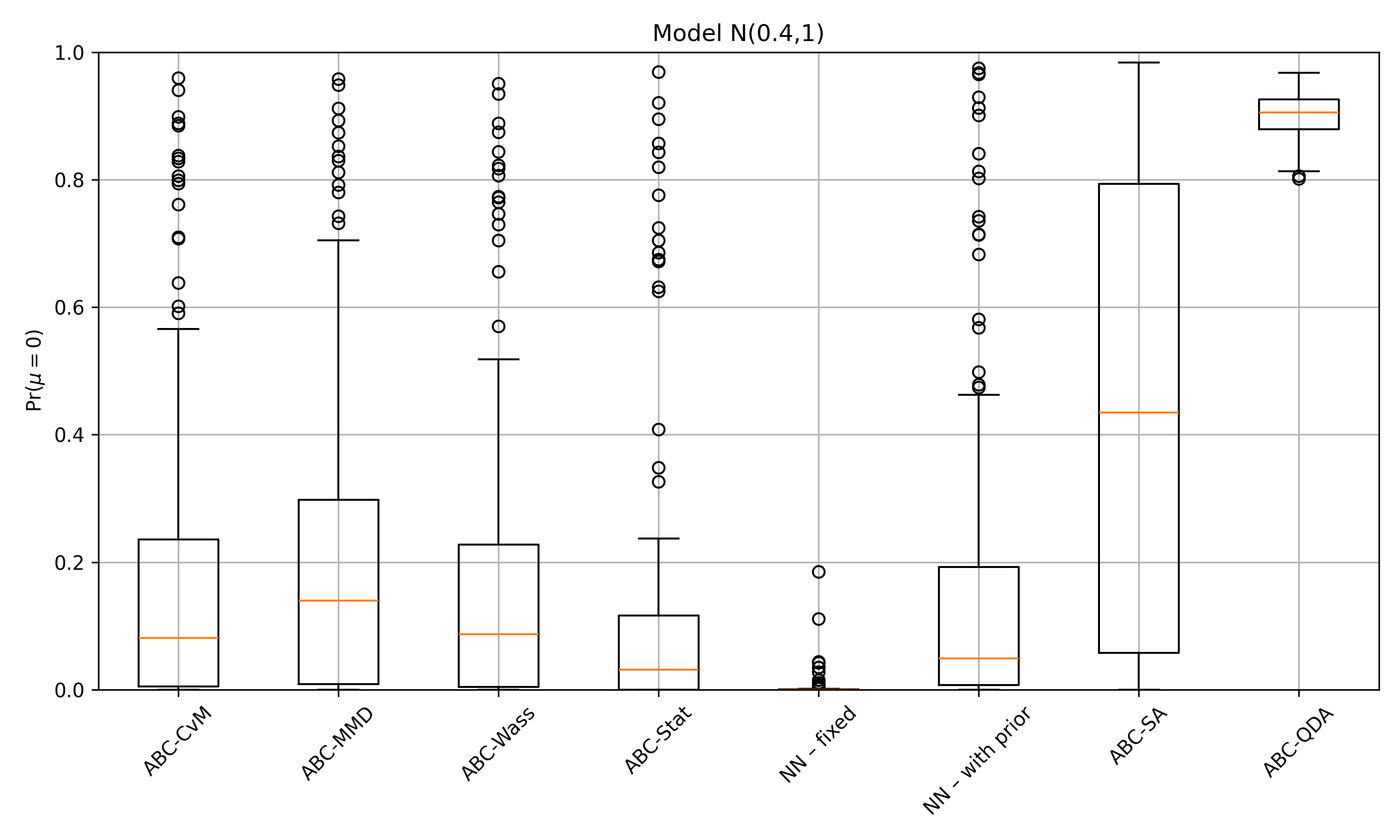}
\caption{Boxplots of the estimated posterior probability of $H_0$ across 100 datasets generated from $N(0.4,1)$.}
\label{fig:normal_boxplots1_4}
\end{figure}
\begin{figure}
\includegraphics[scale=0.5]{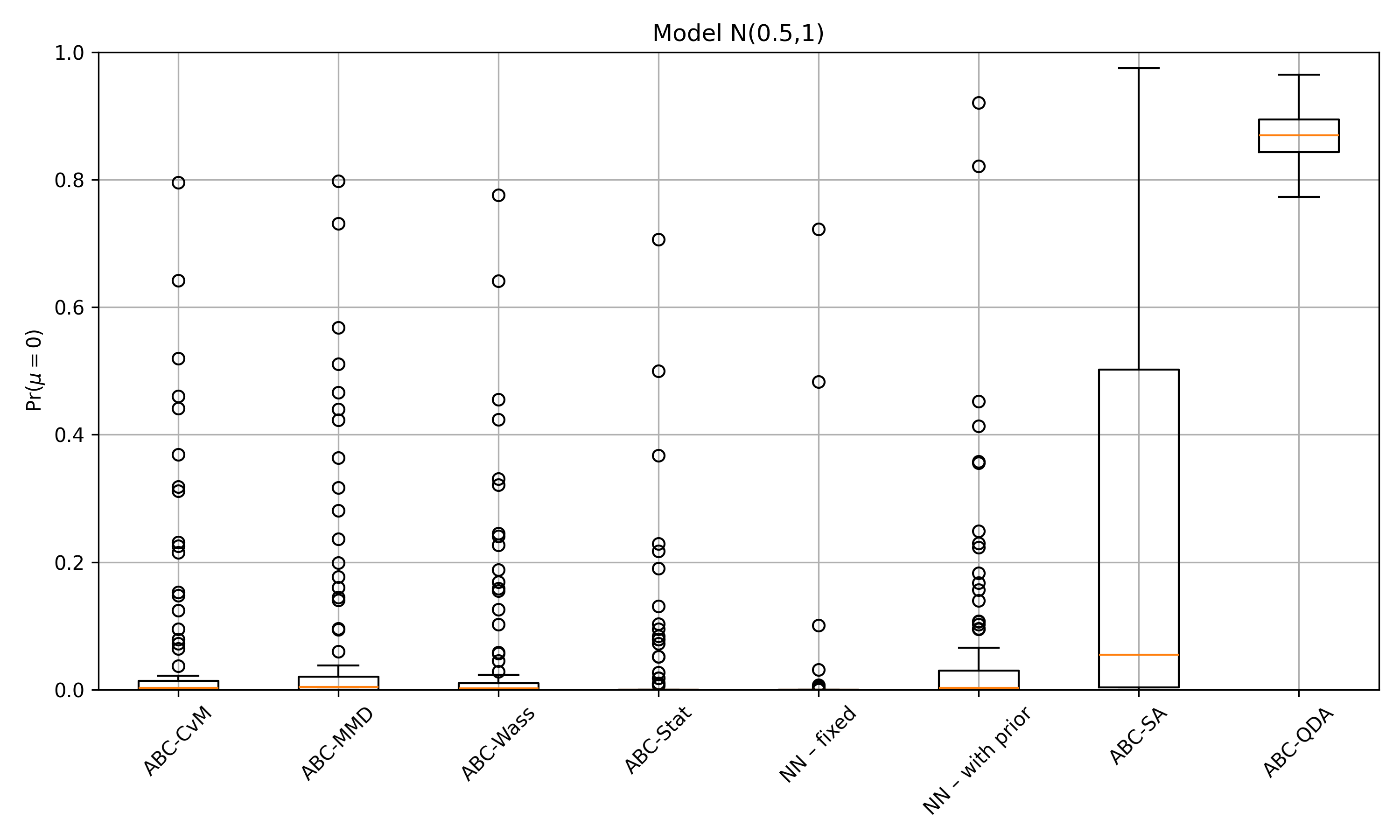}
\caption{Boxplots of the estimated posterior probability of $H_0$ across 100 datasets generated from $N(0.5,1)$.}
\label{fig:normal_boxplots1_5}
\end{figure}

This Appendix presents additional results for Section 4.1 of the main paper, focusing on the case of known variance. Figure \ref{ex_normal_CM} shows the confusion matrices for each problem, while Figures \ref{fig:normal_boxplots0}–\ref{fig:normal_boxplots1_5} display boxplots of the estimated posterior probabilities of selecting $H_0$ across repetitions. Full data ABC methods and ABC-Stat exhibit similar behavior, with high classification accuracy and estimated probabilities of selecting $H_0$ approaching zero as the true mean $\theta_0$ moves further from zero. When $\theta_0$ is small but non-zero, all methods show substantial uncertainty. In comparison, ABC-QDA consistently selects $H_0$ with low uncertainty, and ABC-SA performs better than ABC-QDA and the NN but still shows notable variability in estimated model probabilities. NN with fixed parameters performs well across all settings, displaying the highest uncertainty in the most challenging case ($\theta_0 = 0.1$), but then clearly decreasing the estimated probability of selecting $H_0$ as $\theta_0$ increases. Using prior distributions to generate the parameters strongly increase the performance of NN in this example. 

Results relative to ABC methods are shown using a threshold corresponding to the 0.1\% quantile of distances.

\clearpage
\raggedbottom

\section*{Appendix E: Exponential Family Model Selection}

\begin{figure}[H]
\centering
\includegraphics[scale=0.30]{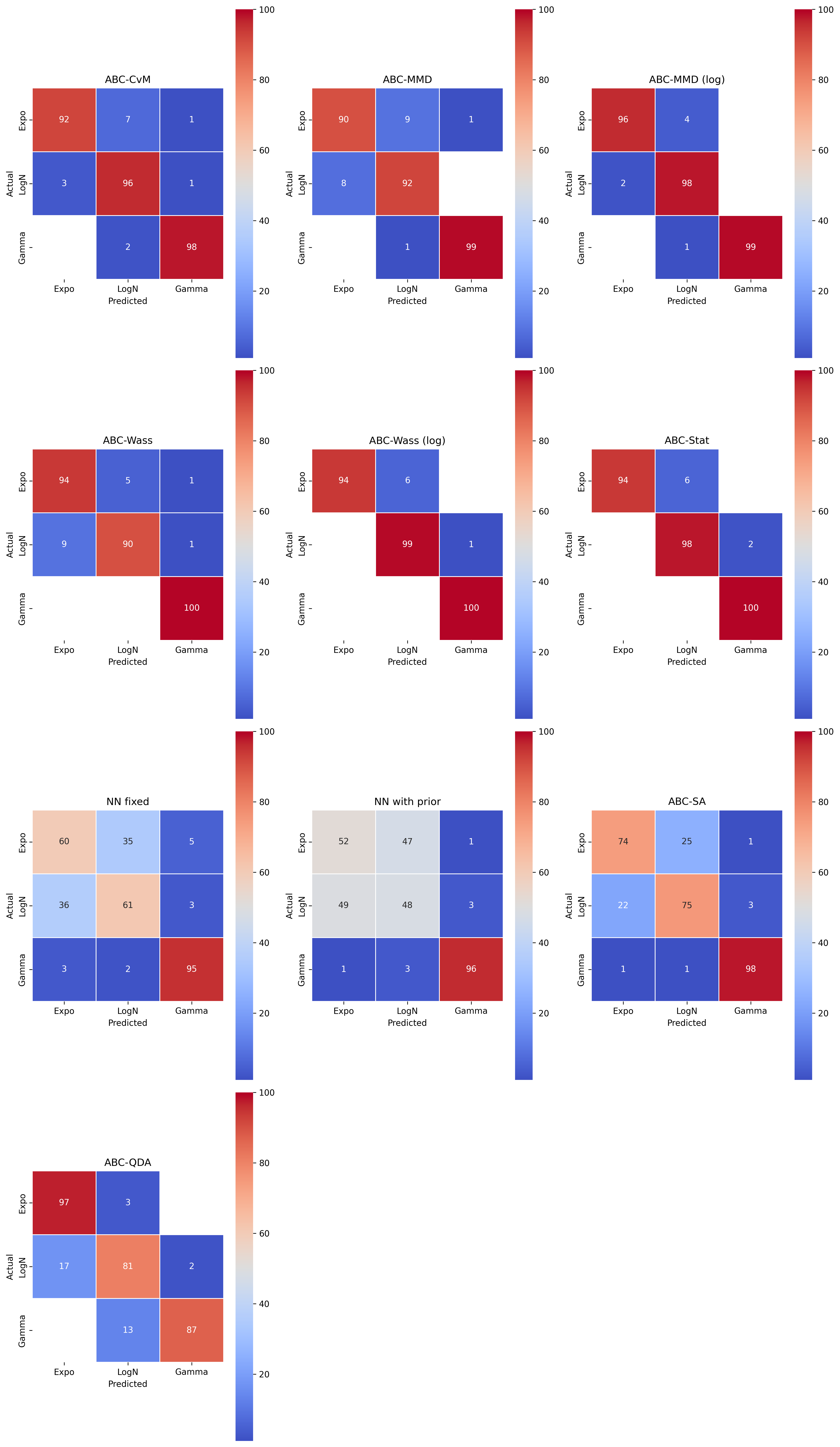}
\caption{Confusion matrices for model selection in the exponential family example with $n=100$.}
\label{ex_expofamily_CM}
\end{figure}

\begin{figure}[H]
\includegraphics[scale=0.5]{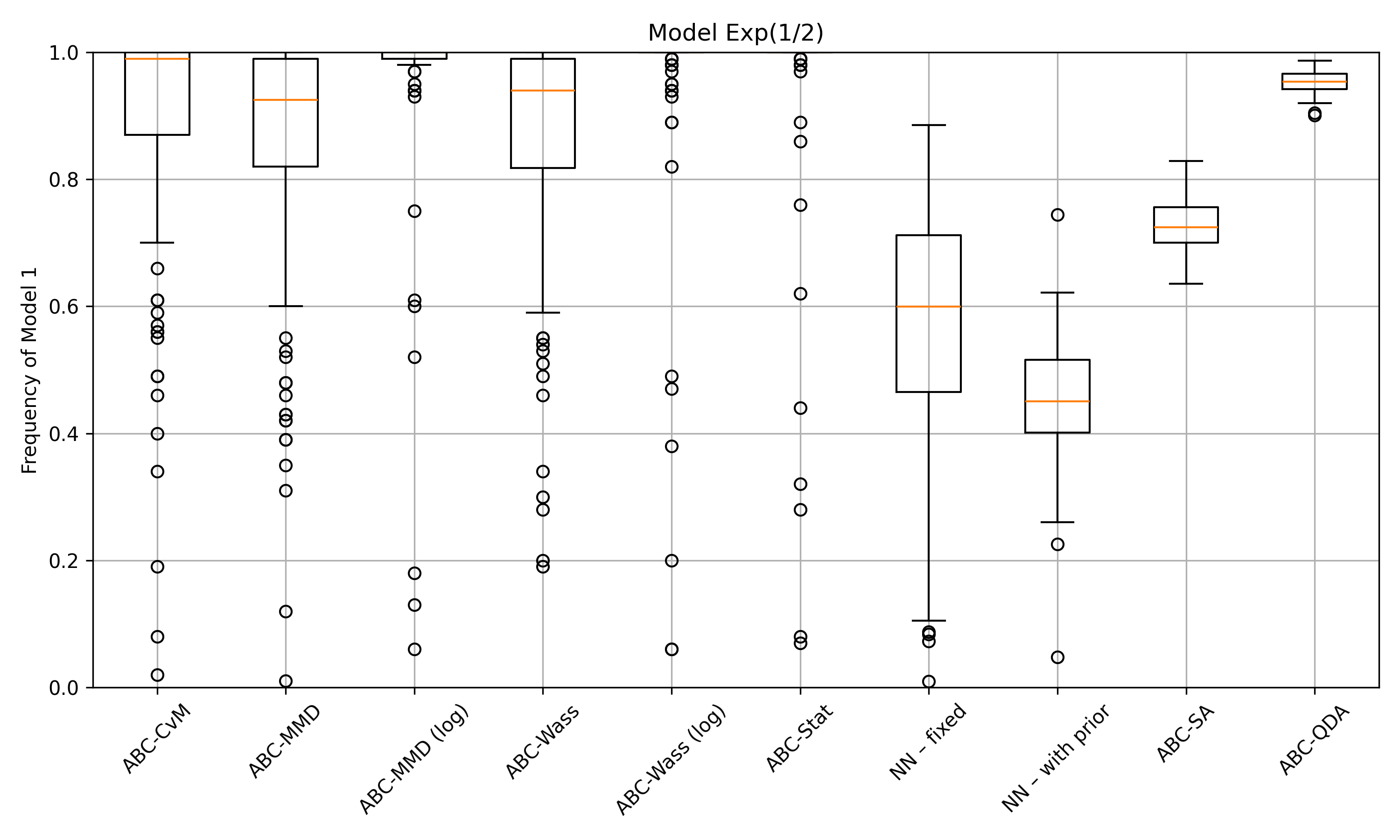}
\caption{Boxplots of the estimated posterior probability of model $M_1$ across 100 datasets generated from $\mathcal{E}\text{xp}(1/2)$.}
\label{fig:expo_family_boxplots_exp}
\end{figure}

\begin{figure}[H]
\includegraphics[scale=0.5]{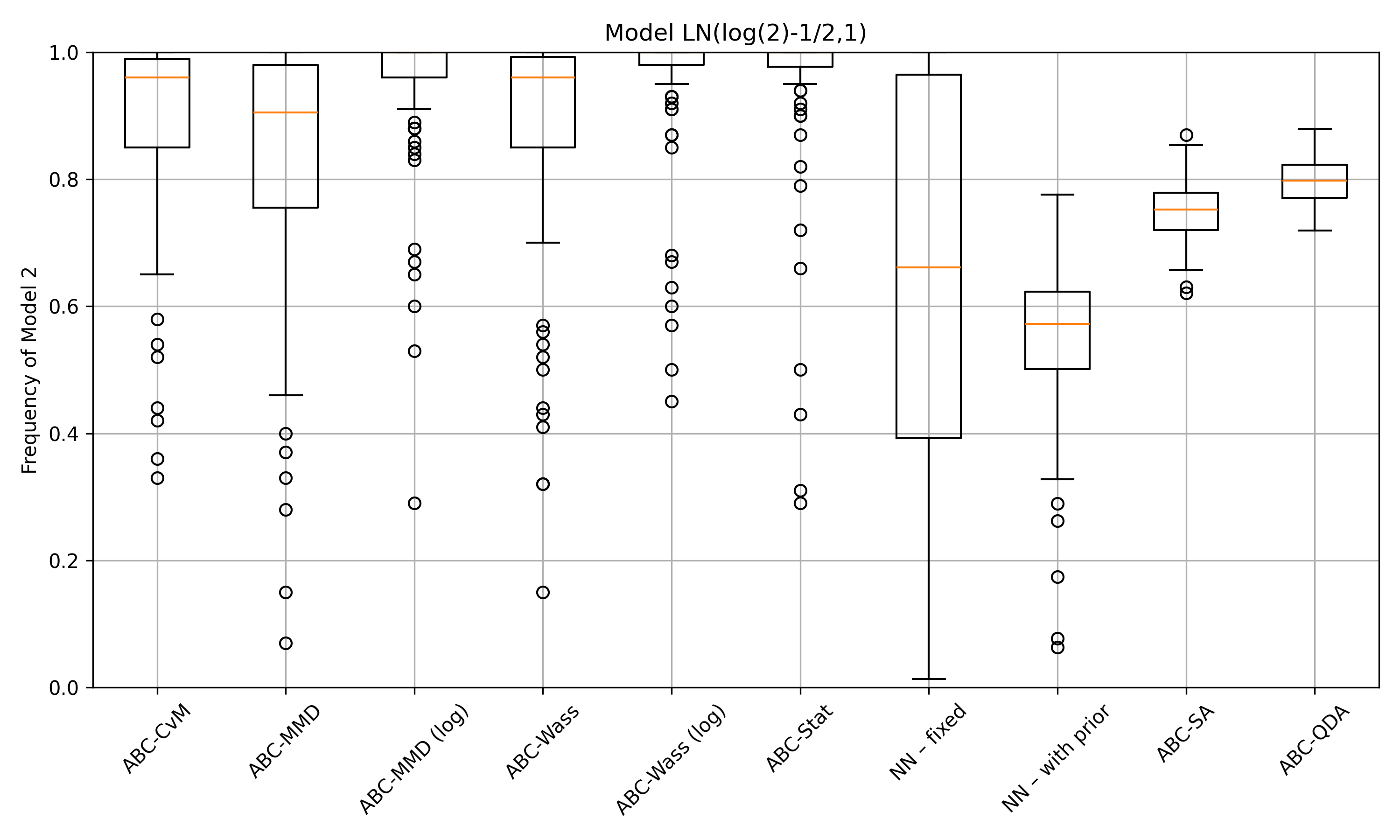}
\caption{Boxplots of the estimated posterior probability of model $M_2$ across 100 datasets generated from $\mathcal{LN}(\log(2)-1/2,1)$.}
\label{fig:expo_family_boxplots_ln}
\end{figure}

\begin{figure}[H]
\includegraphics[scale=0.5]{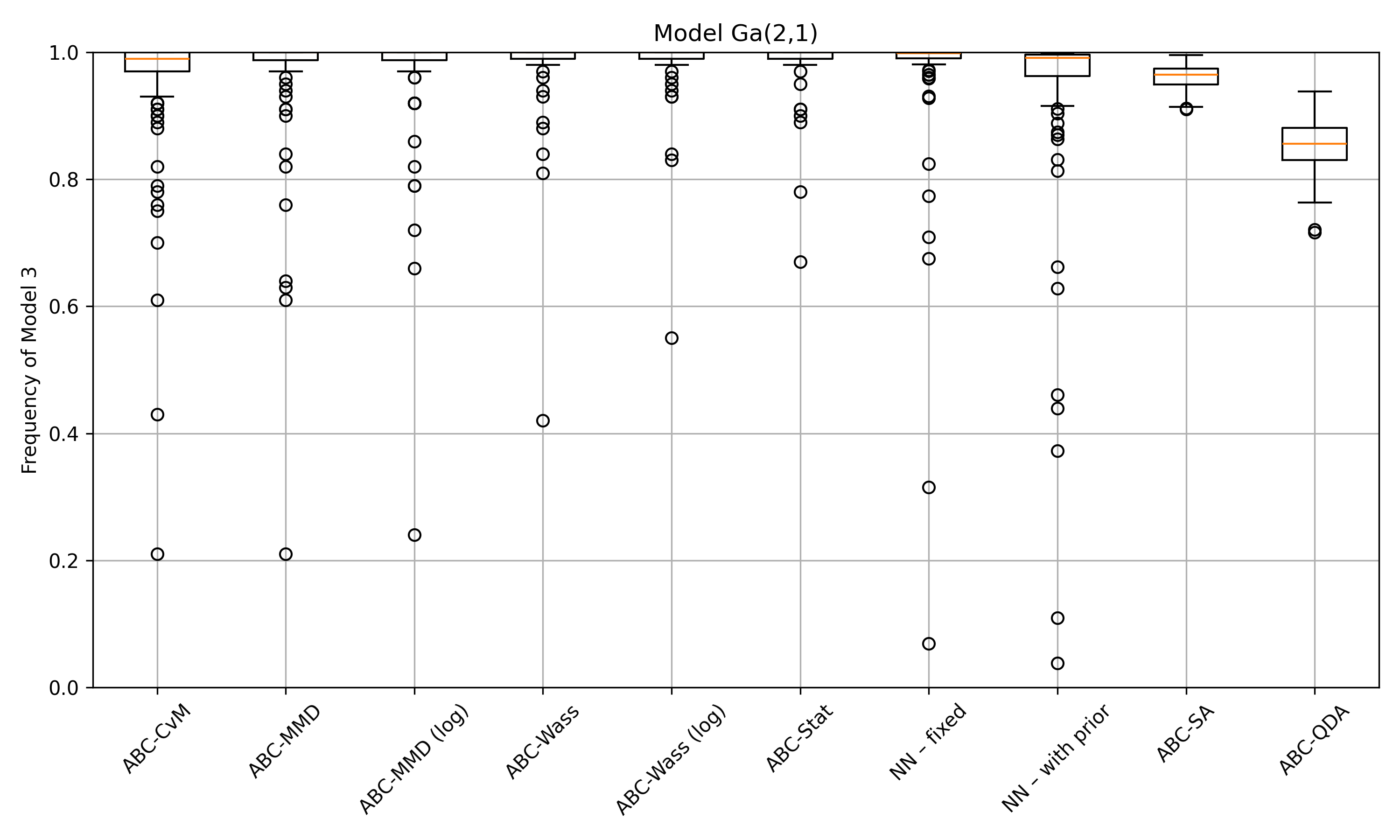}
\caption{Boxplots of the estimated posterior probability of model $M_3$ across 100 datasets generated from $\mathcal{G}\text{a}(2,1)$.}
\label{fig:expo_family_boxplots_gamma}
\end{figure}

Figure~\ref{ex_expofamily_CM} shows the confusion matrices for the exponential family example, confirming the results reported in the main paper: full data ABC approaches achieve near-perfect classification accuracy, closely matching ABC-Stat. In contrast, ABC-SA and ABC-QDA show lower classification performance, while NN (in both the version with fixed parameters and with parameters generated from the prior distributions) displays confusion between the exponential and lognormal models and exhibits the highest variability in the estimated probabilities. The boxplots of the estimated posterior probabilities (Figures~\ref{fig:expo_family_boxplots_exp}–\ref{fig:expo_family_boxplots_gamma}) reveal that all methods display a similar variability. Across methods, the gamma model ($M_3$) is the easiest to identify, with estimated posterior probabilities close to one in most cases. Models $M_1$ and $M_2$ are more challenging to recognize, exhibiting higher variability and less consistent model selection. 

Results relative to ABC methods are shown using a threshold corresponding to the 0.1\% quantile of distances.

\clearpage

\section*{Appendix F: $g$-and-$k$ Distribution Model Selection}

\begin{figure}[h]
    \centering
    \includegraphics[width=8.5cm,height=7.5cm]{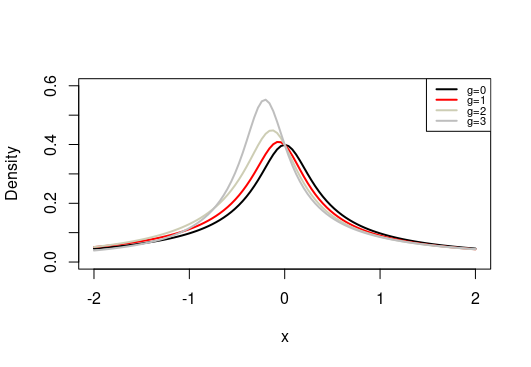}
    \caption{Comparison of $g$-and-$k$ densities under $M_1 \; (g_0 = 0)$ and $M_2 \; (g_0 = 1, 2, 3)$, with $k = 2$ for all models.}
    \label{fig:g-and-k_density}
\end{figure}

Figure~\ref{fig:g-and-k_density} compares densities computed numerically from model $M_1$ ($g=0$) with model $M_2$ for $g=1,2,3$. The cases $g=2$ and $g=3$ are included to highlight how model differences become more pronounced as $g$ increases, whereas for $g=1$ the densities remain close, suggesting that distinguishing between $M_1$ and $M_2$ can be difficult.  Results relative to ABC methods are shown using a threshold corresponding to the 0.1\% quantile of distances.

\begin{figure}[h!]
\centering
\includegraphics[scale=0.25]{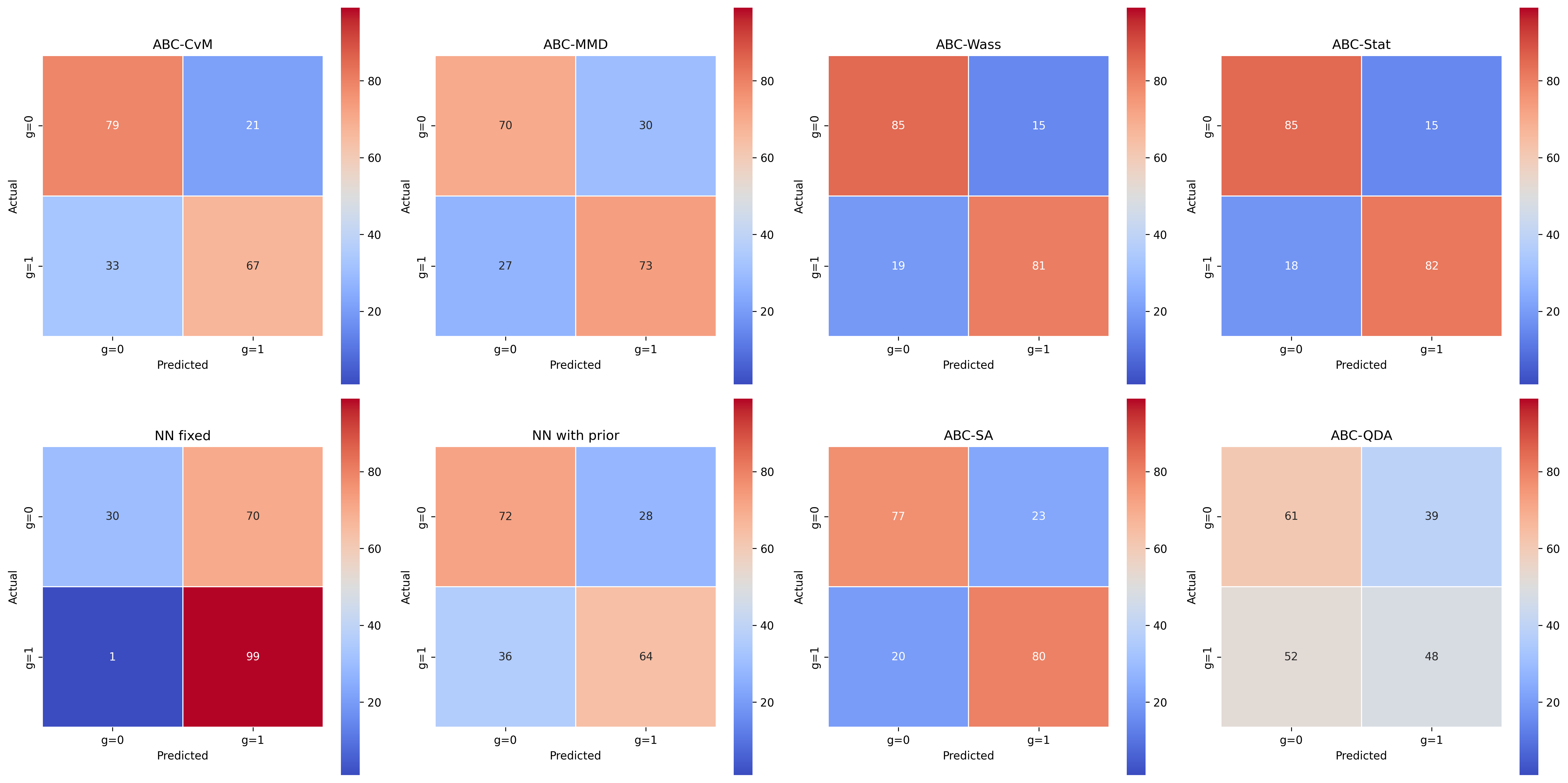}
\caption{Confusion matrices for model selection in the quantile distribution example with $n=100$. ABC methods are shown using a threshold corresponding to the 0.1\% percentile.}
\label{ex_quantile_n100_CM}
\end{figure}

\begin{figure}[h!]
\centering
\includegraphics[scale=0.25]{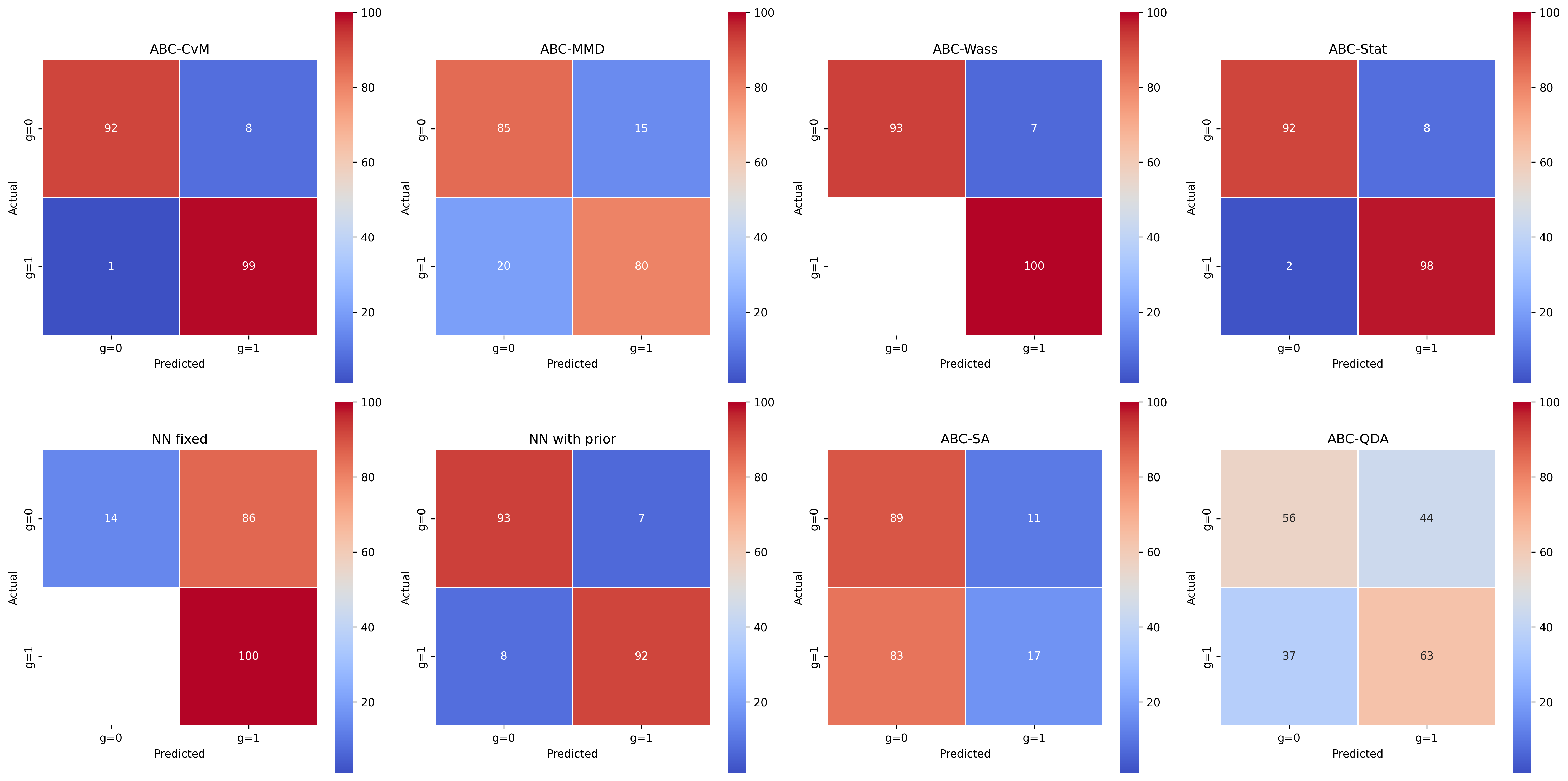}
\caption{Confusion matrices for model selection in the quantile distribution example with $n=1000$. ABC methods are shown using a threshold corresponding to the 0.1\% percentile.}
\label{ex_quantile_n1000_CM}
\end{figure}

The confusion matrices (Figure \ref{ex_quantile_n100_CM} and \ref{ex_quantile_n1000_CM}) highlight that classification is considerably more challenging for samples of size $n=100$ than for $n=1000$. Among the methods, ABC-Wass consistently performs best, achieving perfect classification for data generated under $M_2$ when $n=1000$. By contrast, ABC-QDA appears to struggle with identifying the correct model even with large samples, while NN with fixed parameters tends to favor $M_1$, leading to systematic misclassification when data are generated from $M_0$. Similarly to the example of Section 4.2 and Appendix D (Normal Mean Hypothesis Test), increasing the variability of the parameters and the relative datasets by generating parameters from their prior distributions improves the performance of NN. Within the full data ABC approaches, ABC-MMD shows comparatively weaker performance.

\begin{figure}
\includegraphics[scale=0.5]{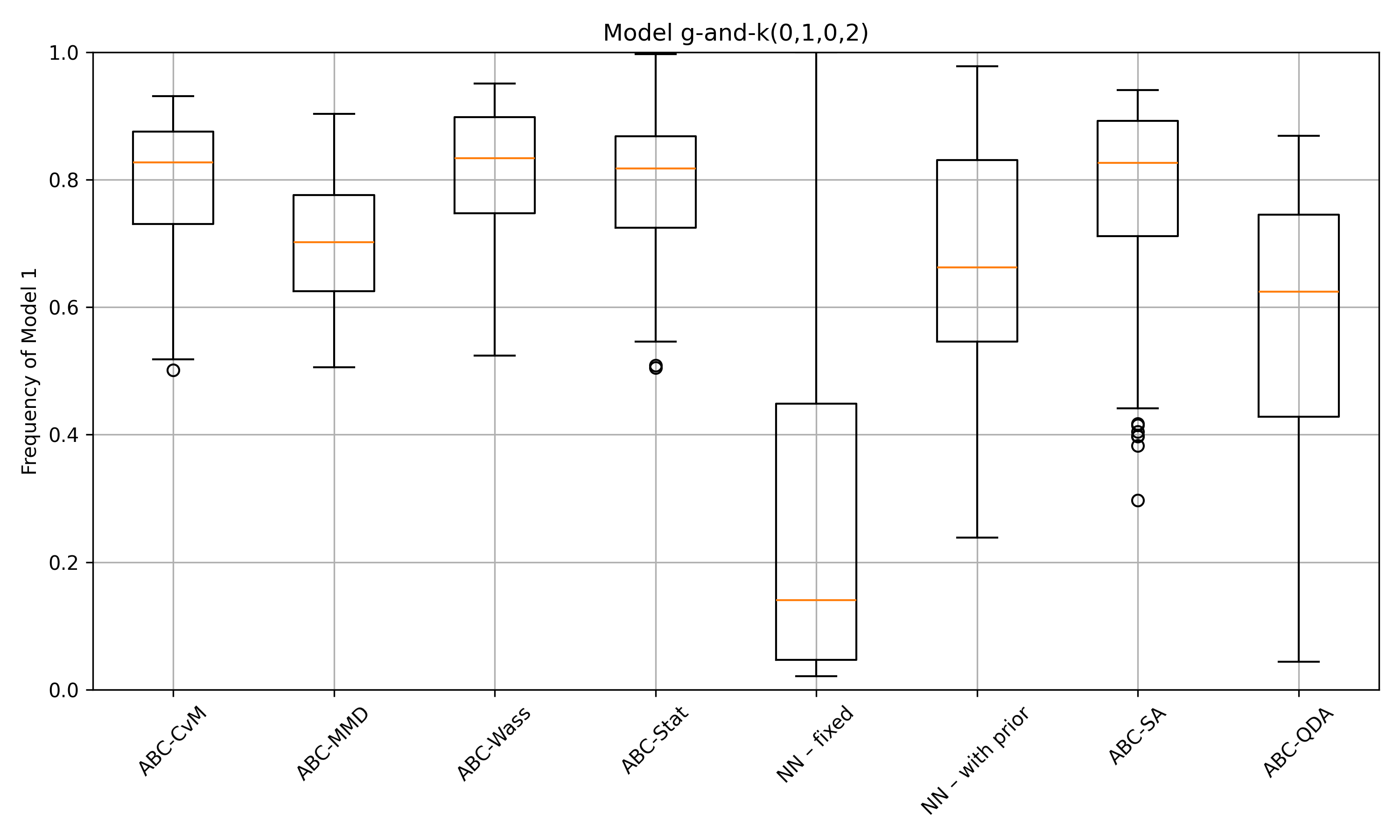}
\caption{Boxplots of the estimated posterior probability of model $M_1$ for 100 datasets generated from a $g$-and-$k$(0,1,0,2) distribution with $n=100$.}
\label{fig:quantile_g0_boxplots}
\end{figure}

\begin{figure}
\includegraphics[scale=0.5]{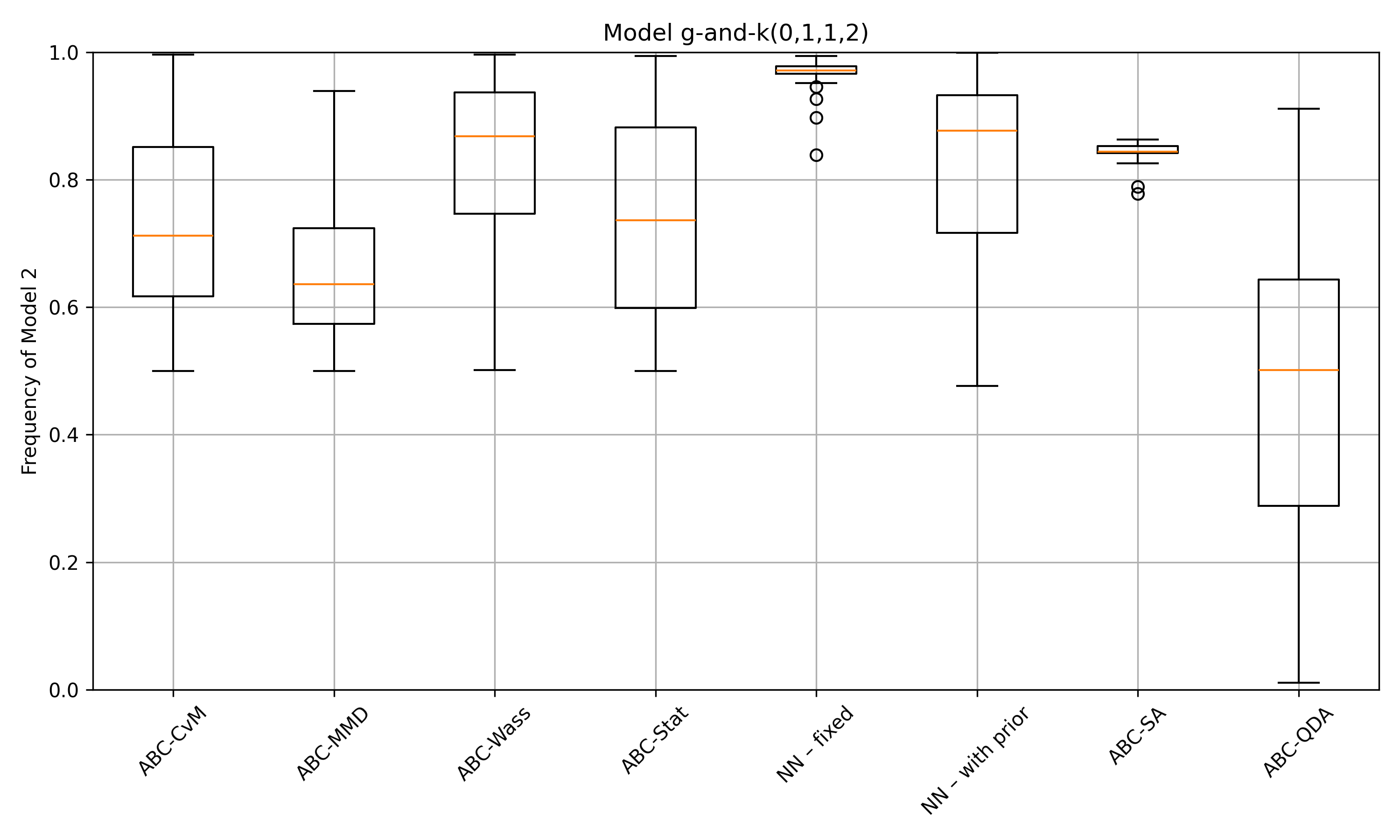}
\caption{Boxplots of the estimated posterior probability of model $M_2$ for 100 datasets generated from a $g$-and-$k$(0,1,1,2) distribution with $n=100$.}
\label{fig:quantile_g1_boxplots}
\end{figure}

\begin{figure}
\includegraphics[scale=0.5]{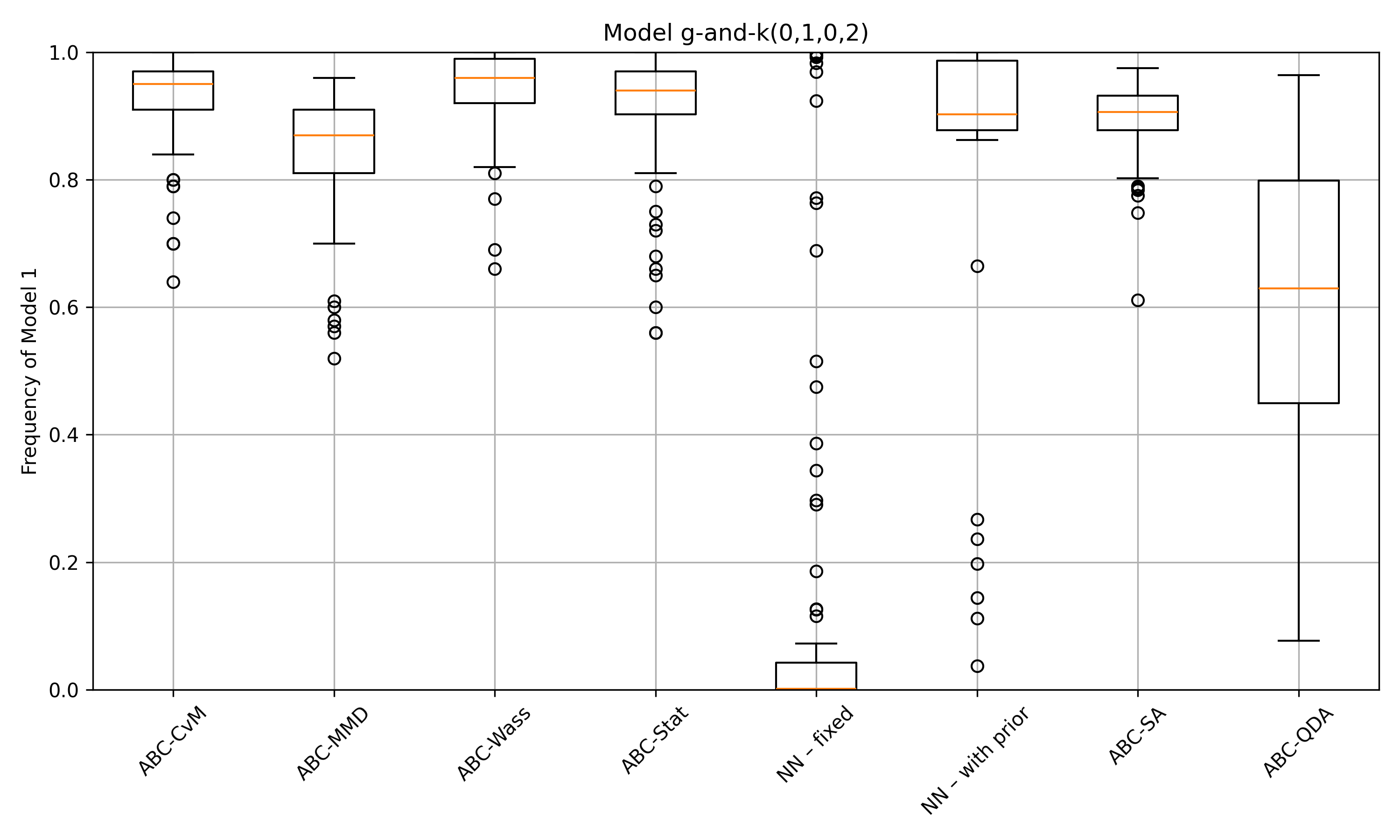}
\caption{Boxplots of the estimated posterior probability of model $M_1$ for 100 datasets generated from a $g$-and-$k$(0,1,0,2) distribution with $n=1000$.}
\label{fig:quantile_g0_n1000_boxplots}
\end{figure}

\begin{figure}
\includegraphics[scale=0.5]{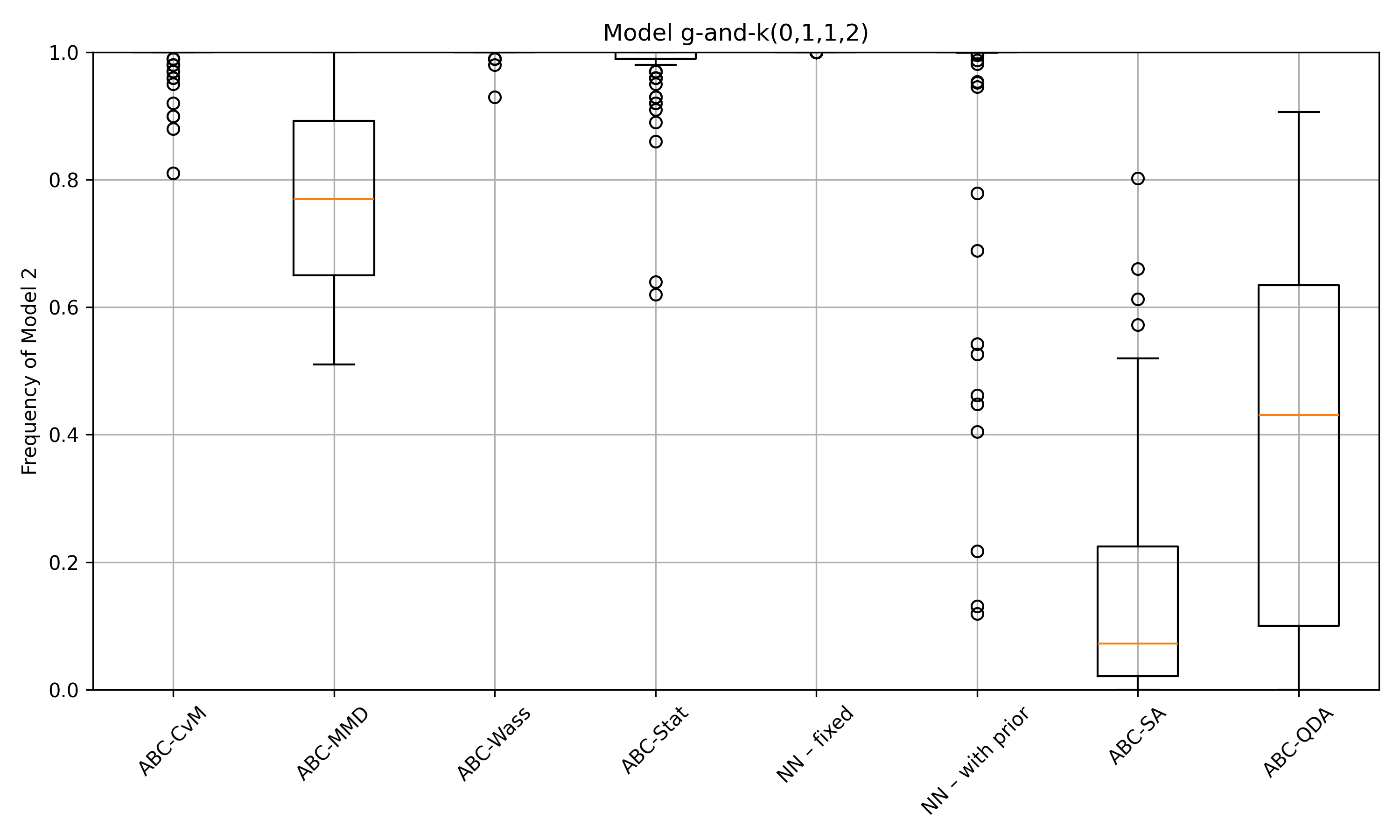}
\caption{Boxplots of the estimated posterior probability of model $M_2$ for 100 datasets generated from a $g$-and-$k$(0,1,1,2) distribution with $n=1000$.}
\label{fig:quantile_g1_n1000_boxplots}
\end{figure}

The boxplots of estimated posterior probabilities of the correct model (Figures \ref{fig:quantile_g0_boxplots}--\ref{fig:quantile_g1_n1000_boxplots}) reveal distinct behaviors across methods. All full data ABC approaches and ABC-Stat exhibit similar performance, with comparable variability and a clear tendency for the estimated probabilities to approach 1 as the sample size increases. ABC-MMD shows the largest variability, particularly when $n=1000$. NN produces posterior probabilities that are typically very close to one when the data are generated from $M_2$, but exhibit substantial variability when the data are generated from $M_1$, reflecting uncertainty and instability in model discrimination. Generating parameters from the prior distribution makes the performance of NN getting closer to the one of ABC approaches. ABC-SA exhibits diminished performance for $n=1000$, likely due to the reduced number of summary statistics used as input to limit memory requirements. Finally, ABC-QDA displays the highest variability among all methods, indicating less stable classification. 

\clearpage

\section*{Appendix G: Toad Movement Model Selection}

\begin{figure}[h!]
\centering
\includegraphics[scale=0.38]{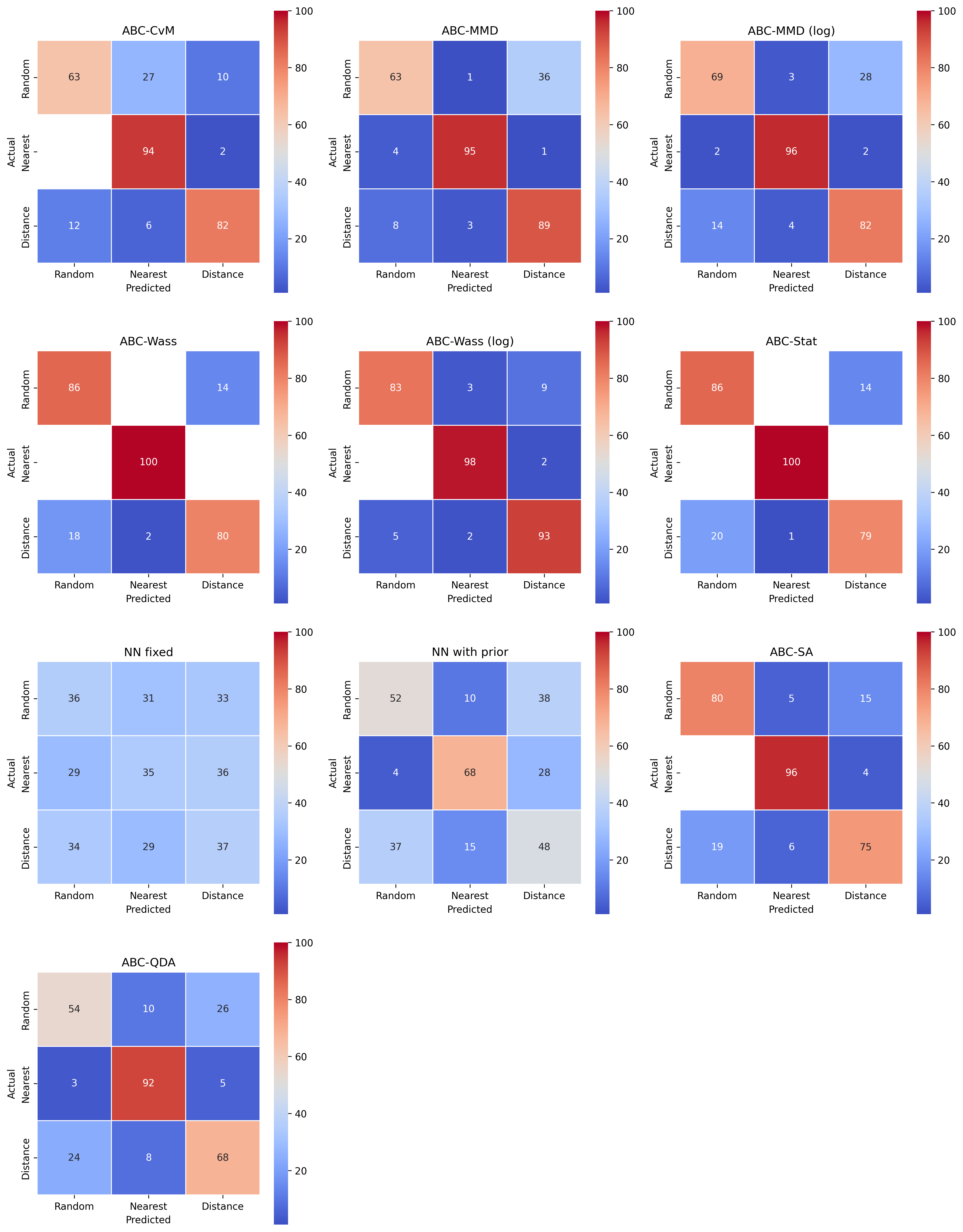}
\caption{Confusion matrices for model selection across methods for the toad movement example.}
\label{ex_toad_CM}
\end{figure}

The confusion matrices for the toad movement model example (Figure \ref{ex_toad_CM}) highlight clear differences in model classification accuracy across methods. ABC-WASS (log) achieves the highest accuracy overall. All full data ABC approaches show nearly perfect classification for model $M_2$, with consistently high but slightly lower accuracy for $M_1$ and $M_3$. Misclassifications occur almost exclusively between $M_1$ and $M_3$, reflecting their greater similarity. NN performs poorly, with accuracies close to random guessing. ABC-SA and ABC-Stat achieve performance comparable to the full data ABC approaches, though with a modest reduction in accuracy. In contrast, ABC-QDA performs the worst among the ABC methods.

\begin{figure}
\includegraphics[scale=0.5]{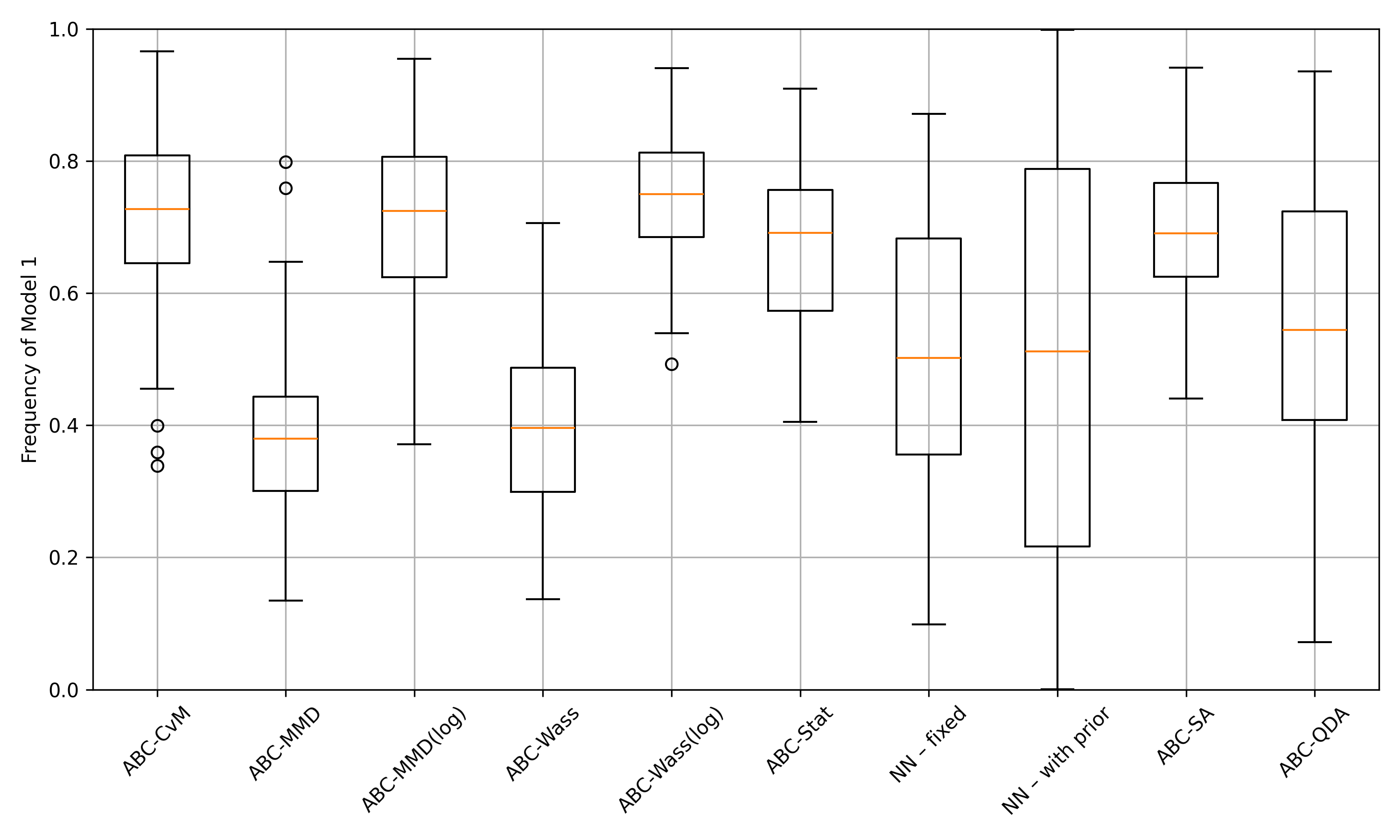}
\caption{Boxplots of the estimated posterior probabilities of model $M_1$ across 100 datasets generated from $M_1$ (random return model).}
\label{fig:toad_random_boxplots}
\end{figure}

\begin{figure}
\includegraphics[scale=0.5]{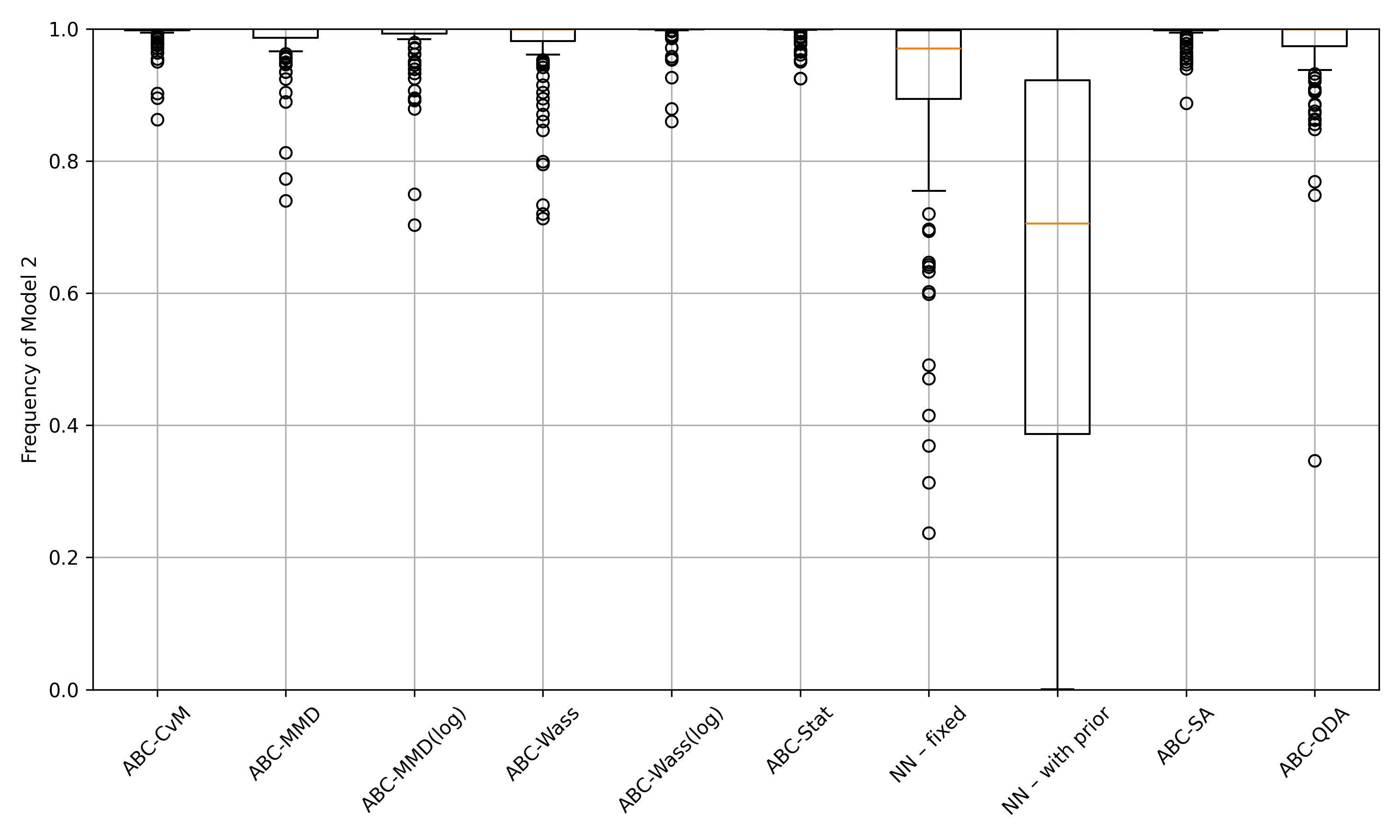}
\caption{Boxplots of the estimated posterior probabilities of model $M_2$ across 100 datasets generated from $M_2$ (nearest return model).}
\label{fig:toad_nearest_boxplots}
\end{figure}

\begin{figure}
\includegraphics[scale=0.5]{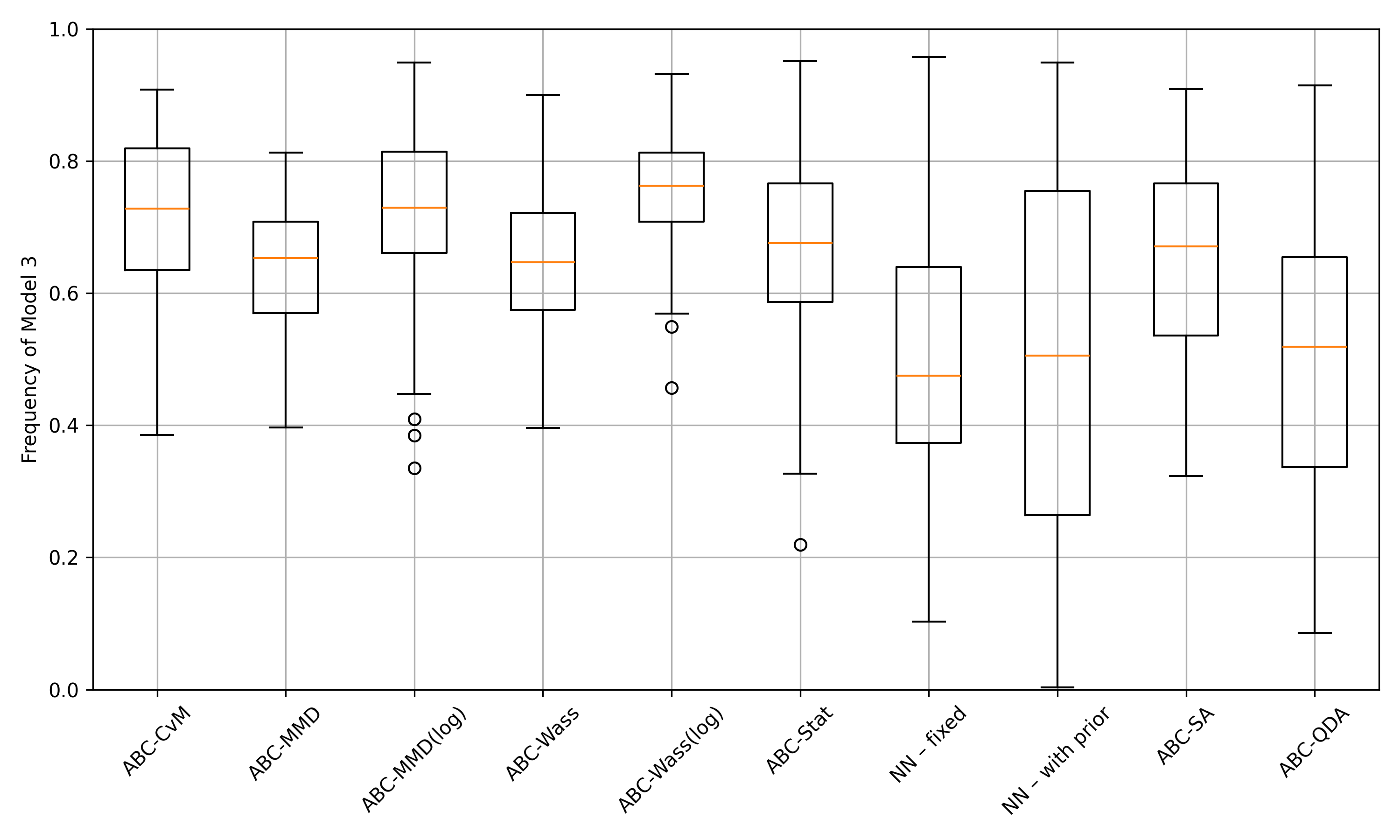}
\caption{Boxplots of the estimated posterior probabilities of model $M_3$ across 100 datasets generated from $M_3$ (distance-based model).}
\label{fig:toad_distance_boxplots}
\end{figure}

Similarly, the boxplots of the posterior probabilities of selecting the correct model (Figures \ref{fig:toad_random_boxplots}-\ref{fig:toad_distance_boxplots}) highlight clear differences between methods. For model $M_2$, most approaches yield probabilities close to one, with the exception of the NN method, which produces many outliers at lower values. In contrast, for models $M_1$ and $M_3$, the variability across methods is larger. The NN and ABC-QDA approaches in particular display the widest spread. Among the ABC methods, ABC-WASS (log) consistently achieves the highest medians with minimal variability, while ABC-WASS and ABC-MMD without log transformation show noticeably reduced accuracy for $M_1$.

Results relative to ABC methods are shown using a threshold corresponding to the 0.1\% quantile of distances.

\end{document}